\documentclass[reprint,aps,prl,showpacs,eqsecnum,twocolumn,superscriptaddress]{revtex4-1}
\usepackage{amsmath,amssymb,graphicx,color,ulem,multirow, inputenc}
\usepackage{hyperref,ulem,url}
\hypersetup{colorlinks=true}

\begin{document}

\title{A magnetar formation in binary neutron star merger} 

\author{Kenta Kiuchi}
\affiliation{Max Planck Institute for Gravitational Physics (Albert Einstein Institute), Am M\"{u}hlenberg, Potsdam-Golm, 14476, Germany}
\affiliation{Center for Gravitational Physics and Quantum Information, Yukawa Institute for Theoretical Physics, Kyoto University, Kyoto 606-8502, Japan}

\author{Alexis Reboul-Salze}
\affiliation{Max Planck Institute for Gravitational Physics (Albert Einstein Institute), Am M\"{u}hlenberg, Potsdam-Golm, 14476, Germany}


\author{Yuichiro Sekiguchi}
\affiliation{Department of Physics, Toho University, Funabashi, Chiba 274-8510, Japan}
\affiliation{Center for Gravitational Physics and Quantum Information, Yukawa Institute for Theoretical Physics, Kyoto University, Kyoto 606-8502, Japan}

\author{Masaru Shibata}
\affiliation{Max Planck Institute for Gravitational Physics (Albert Einstein Institute), Am M\"{u}hlenberg, Potsdam-Golm, 14476, Germany}
\affiliation{Center for Gravitational Physics and Quantum Information, Yukawa Institute for Theoretical Physics, Kyoto University, Kyoto 606-8502, Japan}

\date{\today}

\begin{abstract}
We conduct a global general relativistic neutrino-radiation-transfer magnetohydrodynamics simulation of a $1.35$-$1.35M_\odot$ binary neutron star with the unprecedented spatial resolution of $6.25$\,m on the Japanese supercomputer FUGAKU. The total consumed CPU time is $\approx 530$ million core hours. We initialize the binary neutron star's magnetic field to be $3.16\times 10^{12}$~G at maximum, which is compatible with the upper end of the observed binary pulsars. We demonstrate that the Kelvin-Helmholtz instability that emerges when the two neutron stars touch amplifies the magnetic field to an expected electromagnetic saturation energy of $\sim 10^{50}$~erg within $3$~ms after the merger. The spectral analysis indicates that the Kazantsev and Kolmogorov spectra are reproduced in the magnetic and kinetic power spectral densities, respectively. We also find that it induces stellar-scale magnetic field amplification by at least a factor of $316$. We conclude that a magnetar may form at least temporarily following neutron star mergers in a few ms. 
\end{abstract}

\maketitle

{\it Introduction}.--
Multimessenger observations GW170817, GRB170817A, and AT2017gfo proved that a binary neutron star merger is one of the most interesting sources of relativistic astrophysical transients~\cite{LIGOScientific:2017vwq,LIGOScientific:2017ync,LIGOScientific:2020aai,LIGOScientific:2018hze,LIGOScientific:2018cki,LIGOScientific:2017ync}. 
Since their detections, tremendous effort has been made to understand these events theoretically/observationally~\cite{Arcavi:2017xiz,Alexander:2017aly,Bauswein:2017vtn,Banerjee:2026paw,Banerjee:2022doa,Banerjee:2020myd,Breschi:2021tbm,Coughlin:2018fis,Coughlin:2018miv,Chiba:2026flt,Chornock:2017sdf,Coulter:2017wya,Cowperthwaite:2017dyu,Christie:2019lim,Decoene:2019eux,Dietrich:2020efo,Domoto:2024qfp,Domoto:2023lrv,Domoto:2022cqp,Domoto:2021xfq,Drout:2017ijr,De:2018uhw,DES:2017kbs,Espino:2023mda,Espino:2023dei,Farrar:2024zsm,Farrar:2025kpk,Fernandez:2018kax,Fong:2017ekk,Fujibayashi:2017puw,Fujibayashi:2020jfr,Fujibayashi:2020dvr,Fujibayashi:2020qda,Fujibayashi:2022ftg,Foucart:2018lhe,Foucart:2024npn,Foucart:2024kci,Foucart:2022kon,Foucart:2021ikp,Gottlieb:2023sja,Gottlieb:2024mwu,Hajela:2019mjy,Hajela:2021faz,Hamidani:2019qyx,Hamidani:2023wky,Hamidani:2023gzj,Hotokezaka:2018,Hotokezaka:2023aiq,Hotokezaka:2022rci,Hotokezaka:2021ofe,Hammond:2022uua,Hammond:2021vtv,Ishizaki:2021,Ishizaki:2021vsx,Ioka:2017nzl,Jacobi:2025eak,Kiuchi:2017zzg,Kiuchi:2022nin,Kiuchi:2023obe,Kato:2024jke,Kawaguchi:2020vbf,Kawaguchi:2019nju,Kawaguchi:2018ptg,Kitoviene:2025cdf,Kasen:2017sxr,Kasliwal:2017ngb,Kilpatrick:2017mhz,Kawaguchi:2018ptg,Lippuner:2017bfm,Musolino:2024sju,Musolino:2024xqy,Most:2018eaw,Most:2019kfe,Most:2020bba,Most:2021ktk,Most:2022yhe,Most:2023sft,Most:2023sme,Most:2025kqf,Mosta:2020hlh,Milton:2018,Margalit:2017dij,Margalit:2019dpi,Margutti:2018xqd,Metzger:2018qfl,Mooley:2018qfh,McCully:2017lgx,Metzger:2021grk,Margutti:2017cjl,Neuweiler:2024jae,Nedora:2020hxc,Perego:2017wtu,Perego:2019adq,Pais:2022ynf,Pais:2024mpw,Pognan:2024ise,Pognan:2022pix,Pognan:2023qhw,Pognan:2025eat,Pognan:2021wpy,Radice:2017lry,Radice:2023zlw,Radice:2020ddv,Radice:2018a,Radice:2018ghv,Radice:2018xqa,Rossi:2019fnm,Rosswog:2024vfe,Sadeh:2024ywm,Savchenko:2017ffs,Shappee:2017zly,Smartt:2017fuw,Shibata:2017xdx,Shibata:2019ctb,Shibata:2021bbj,Shibata:2021xmo,Shibata:2017xht,Sridhar:2020uez,Shrestha:2023exe,Tanaka:2017qxj,Tanaka:2019iqp,Tarumi:2023apl,Tanvir:2017pws,Vieira:2022tnm,Villar:2017wcc,Wanajo:2018wra,Wanajo:2021jzd,Waxman:2017sqv,Zappa:2022}. 
However, the complete picture has not yet been drawn. 
The key challenge on the theoretical side is to conduct ab initio, high-resolution, long-term, global numerical relativity simulations for binary neutron star mergers, in which the effects of all fundamental interactions are implemented in a self-consistent manner~\cite{Kiuchi:2022nin,Kiuchi:2023obe,Hayashi:2021oxy}. Even with the currently available computational resources, it remains challenging, and previous works suffer from limitations, such as the need for unrealistically large magnetic-field initialization~\cite{Kiuchi:2014,Kiuchi:2017zzg,Kiuchi:2022nin,Kiuchi:2023obe,Musolino:2024sju}, the need to stick with a semi-global simulation~\cite{Gutierrez:2026ngt}, or the need to implement a sub-grid model~\cite{Aguilera-Miret:2020dhz,Aguilera-Miret:2021fre,Aguilera-Miret:2023qih,Aguilera-Miret:2024cor,Aguilera-Miret:2025nts,Palenzuela:2021gdo,Most:2023sft,Most:2023sme}. 

Particularly, it is an open question how the neutron star's magnetic field becomes dynamically important~\cite{Kiuchi:2025}, which is likely to be necessary for launching a relativistic jet to explain GRB170817A~\cite{Kiuchi:2023obe} and for driving the large amount of neutron-rich matter ejection enough to explain AT2017gfo~\cite{Kiuchi:2023obe}. The Kelvin-Helmholtz instability is one of the promising mechanisms for amplifying the neutron star's magnetic field on a very short timescale to the magnetar's strength~\cite{Rasio:1999,Price:2006,Kiuchi:2015,Tripathi:2026}. Note that the growth rate of this instability is proportional to the wavenumber, and this instability activates only for a couple of milliseconds after the merger~\cite{Kiuchi:2025}. 
More importantly, the generation of the large-scale (stellar-scale) field via this instability remains uncertain because it primarily produces the small-scale turbulent field. Therefore, one needs to conduct ultra-high-spatial-resolution global numerical-relativity magnetohydrodynamics simulations by initializing the neutron star's magnetic field to be compatible with binary pulsar observations~\cite{Lorimer:2008se}. 

In this Letter, we report the results of a global numerical-relativity neutrino-radiation-transfer magnetohydrodynamics simulation of a binary neutron star with $6.25$\,m spatial resolution, conducted on the Japanese supercomputer FUGAKU. 

{\it Numerics, model, and grid setup}.--
We employ the numerical relativity code {\tt NANASI}~\cite{Kiuchi:2022}. {\tt NANASI} implements the Baumgarte-Shapiro-Shibata-Nakamura-puncture formulation together with the Z4c constraint propagation prescription to solve Einstein's equation~\cite{Shibata:1995,Baumgarte:1998te,Baker:2005vv,Campanelli:2005dd,Hilditch:2012fp}. Fourth-order finite differencing is employed for the spatial derivative and fourth-order Runge-Kutta for the time integration of the metric. It also implements the general relativistic second-order Harten-Lax-van Leer-discontinuities finite volume Riemann solver and the Gardiner-Stone constrained transport solver together with the third-order picewise-parabolic-method cell reconstruction for the relativistic magnetohydrodynamics~\cite{Mignone:2005ft,Gardiner:2007nc,White:2015omx,1984JCoPh..54..174C}. For the neutrino radiation transfer, we employ the gray M1+GR Leakage transport scheme to take into account the heating and cooling~\cite{Sekiguchi:2010ep,Sekiguchi:2012uc}. 

{\tt NANASI} employs the nested grid structure based on the Berger-Oliger mesh refinement algorithm to cover a wide dynamic range~\cite{Berger:1984zza}. Specifically, the 2:1 refinement rule and the adaptive time-marching algorithm are employed. 

As a binary neutron star model, we assume DD2 as the nuclear equation of state~\cite{Hempel:2009mc} and a $1.35$-$1.35M_\odot$ equal-mass irrotational binary. The nuclear equation of state is stitched with the Helmholtz equation of state~\cite{Timmes} to extend to the low-density and temperature region~\cite{Hayashi:2021oxy}. 
An initial orbital separation is $\approx 44.4~{\rm km}$. A public spectral library {\tt FUKA} is used to generate the orbital eccentricity-reduced initial data~\cite{Papenfort:2011}. 

Following Ref.~\cite{Kiuchi:2023obe}, we initialize the magnetic field by the toroidal component of the vector potential as
\begin{align}
A_\varphi \propto \max\left(P-0.0004 P_{\rm max},0\right)^2, \label{eq:vector}
\end{align}
where $P$ and $P_{\rm max}$ denote the pressure and its maximum. The overall normalization is set to be the initial maximum field strength $B_{0,{\rm max}}=3.16 \times 10^{12}~{\rm G}$, which is compatible with the observed binary pulsars~\cite{Lorimer:2008se}. We should note that the dipole field strength inferred from the observation could be located at the pole, while that with Eq.~(\ref{eq:vector}) is located inside the neutron star. However, a free precession observed in the magnetar indicates that the interior magnetic field strength could be stronger than the surface magnetic field~\cite{Makishima:2014dua}. Therefore, the field strength assumed in this Letter is a conservative choice. 

The nested grid is composed of the concentric Cartesian domain with the grid number $N$ and the grid spacing $\Delta x_{\rm (lv)}$ for all three spatial coordinates. The size of a nested domain {\rm lv} is $L_{\rm (lv)}=[(-N-1/2)\Delta x_{\rm (lv)}:(N+1/2)\Delta x_{\rm (lv)}]$ in one spatial  direction. The orbital plane symmetry is assumed. We employ $16$ nested domains with $\Delta x_{(16)}=6.25$\,m and $N=716$. With this, the central region of $L_{(16)}\approx 9$~km, where the Kelvin-Helmholtz instability sets in, is covered by the finest resolution, while the entire binary neutron star is covered by the domain $13$ with $\Delta x_{(13)}=50$~m. {\tt NANASI} employs the conservative mesh refinement which preserves the baryonic mass within $O(10^{-6}$--$10^{-5})\%$ accuracy and both the divergence-free condition and magnetic flux conservation with the machine precision, respectively~\cite{Kiuchi:2012qv,Kiuchi:2022,Neuweiler:2024jae}. 
The atmospheric density and temperature are set to be $0.1667~{\rm g~cm^{-3}}$ and $10^{-3}$~MeV, respectively. We use $\approx 4.4 \times 10^5$ cores on FUGAKU. 

\begin{figure*}[t]
 	 \includegraphics[width=0.4\linewidth]{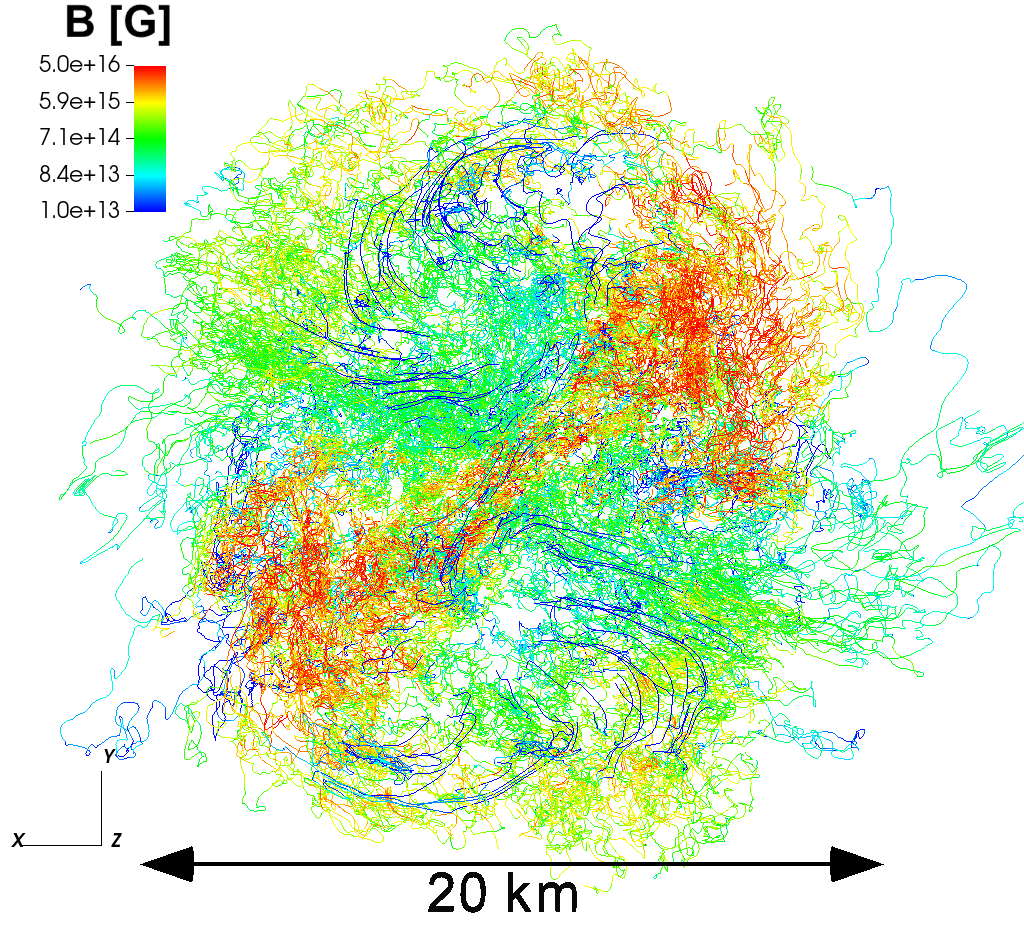}    
      \includegraphics[width=0.4\linewidth]{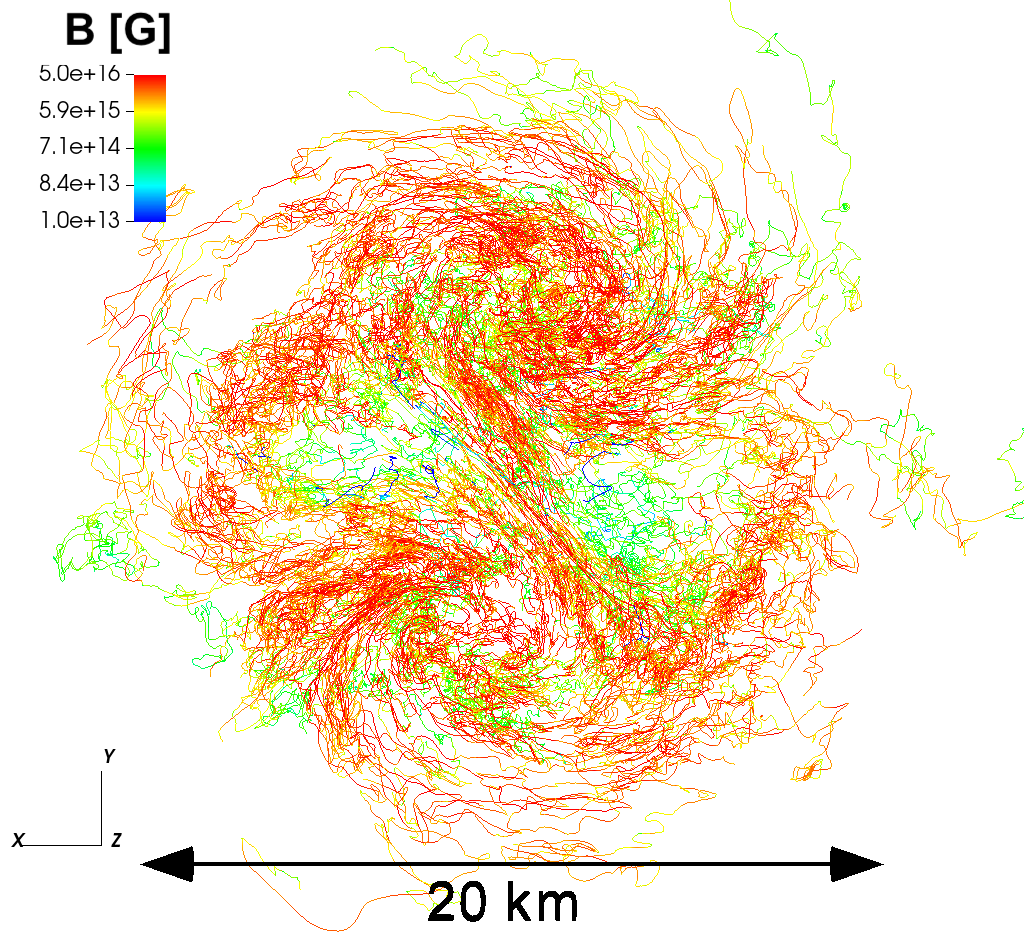}  	 
 	 \caption{The magnetic field lines at $t-t_\mathrm{merger}\approx 2$~ms (left) and at $5$~ms (right) seen from below the equatorial plane. The 3D visualization is in \cite{3D-viz.}.
      }\label{fig:B-field-lines}
\end{figure*}

\begin{figure}[t]
 \includegraphics[width=0.99\linewidth]{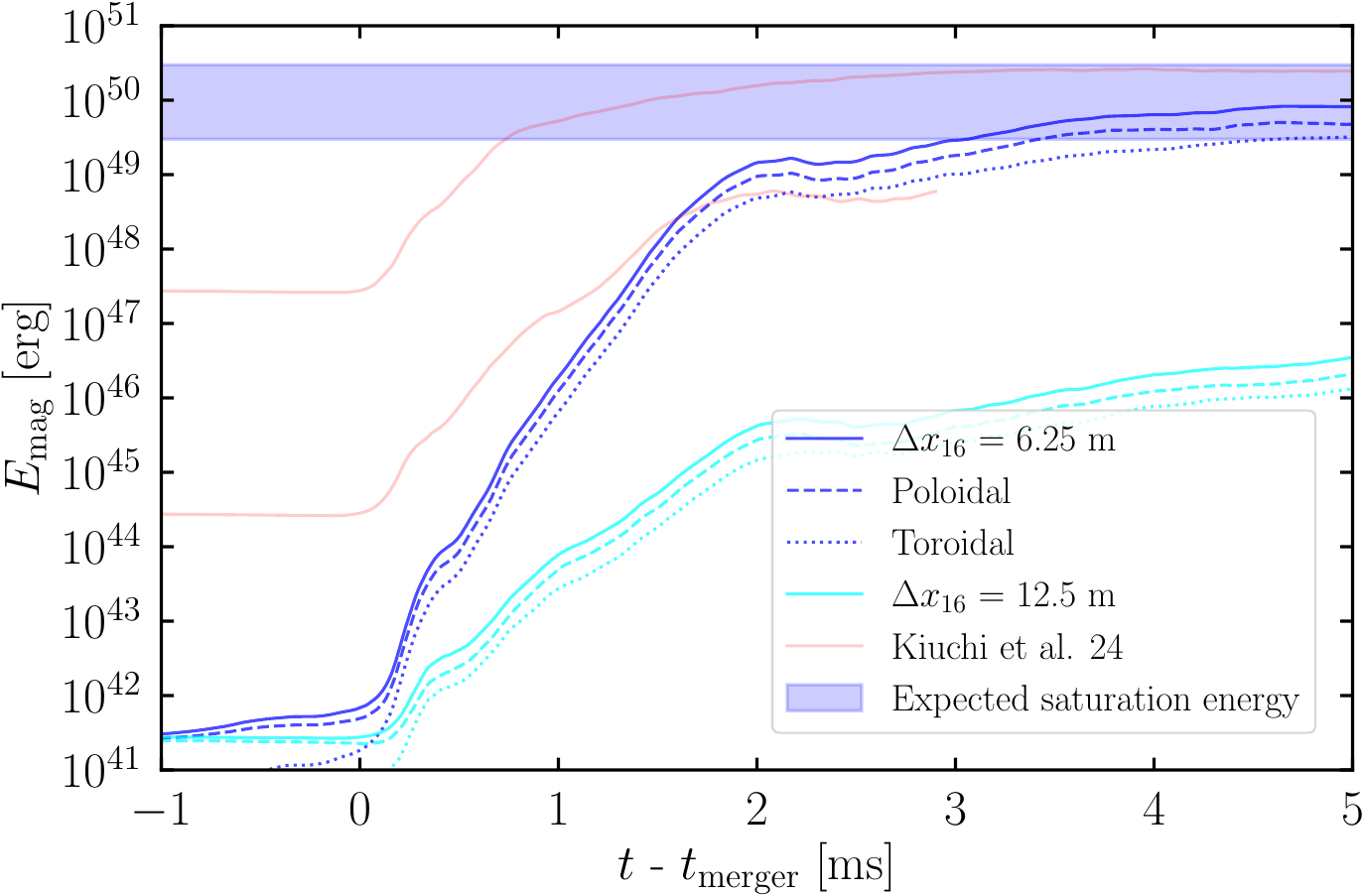}              
 \caption{Electromagnetic energy as a function of the post-merger time. The blue-solid, dashed, and dotted curves show the total, poloidal, and toroidal-field energy, respectively. The blue-shaded region is an expected saturation energy due to the Kelvin-Helmholtz instability~\cite{Aguilera-Miret:2020dhz,Aguilera-Miret:2021fre,Aguilera-Miret:2023qih,Aguilera-Miret:2024cor,Aguilera-Miret:2025nts,Palenzuela:2021gdo,Gutierrez:2026ngt,Kiuchi:2023obe}. The thin red curves plot our previous simulations~\cite{Kiuchi:2023obe}. 
      }\label{fig:KH}
 \end{figure}

{\it Result}.--
Figure~\ref{fig:B-field-lines} plots the magnetic-field lines at $t-t_\mathrm{merger}\approx 2~{\rm  ms}$ and $5~{\rm ms}$. We define the merger time as the time of the gravitational wave peak amplitude. 
It suggests the magnetic field with $\approx 10^{16}~{\rm G}$ (locally, $10^{17}~{\rm G}$) is produced. 
Figure~\ref{fig:KH} plots the electromagnetic energy as a function of the post-merger time. 
The blue-solid, dashed, and dotted curves denote the total, poloidal, and toroidal electromagnetic energy, and both the poloidal and toroidal components exhibit clear exponential growth. The thin red curves show our previous simulation data, in which we employed the same binary neutron star model, but we initialized the magnetic field to be $10^{14}{~\rm G}$ and $10^{15.5}~{\rm G}$, and ran the simulation with $\Delta x_{(16)}=12.5$~m~\cite{Kiuchi:2023obe}. We also run a low resolution simulation with $\Delta x_{(16)}=12.5$~m, $N=361$, and $B_\mathrm{0,max}=3.16\times 10^{12}$~G (the cyan curves). 
The simulation with $\Delta x_{(16)}=6.25\,{\rm m}$ results in the larger growth rate than the simulations with $\Delta x_{(16)}=12.5~{\rm m}$, which is the natural consequence of the Kelvin-Helmholtz instability~\cite{Rasio:1999,Price:2006,Kiuchi:2014,Kiuchi:2017zzg,Kiuchi:2023obe,Aguilera-Miret:2020dhz,Aguilera-Miret:2021fre,Aguilera-Miret:2023qih,Aguilera-Miret:2024cor,Aguilera-Miret:2025nts,Palenzuela:2021gdo,Gutierrez:2026ngt}.

A blue-shaded region is the expected electromagnetic saturation energy via the Kelvin-Helmholtz instability predicted by the high-resolution global simulation~\cite{Kiuchi:2023obe}, and the global simulations implementing the gradient sub-grid scale model~\cite{Aguilera-Miret:2020dhz,Aguilera-Miret:2021fre,Aguilera-Miret:2023qih, Aguilera-Miret:2024cor,Aguilera-Miret:2025nts,Palenzuela:2021gdo,Gutierrez:2026ngt}. 
At the end of the simulation of $t-t_{\rm merger}\approx 5$~ms, the electromagnetic energy reaches the expected saturation energy. In this particular model, the rapid growth of the magnetic fields by the Kelvin-Helmholtz instability ends at $t-t_{\rm merger} \approx 2$~ms due to the dissipation of the shear layer by the shock waves~\cite{Kiuchi:2023obe}. Interestingly, the electromagnetic energy continues to be amplified even after the Kelvin-Helmholtz instability phase. 

To understand the prominent growth of the magnetic field more precisely, we conduct the power spectral density (PSD) analysis of the electromagnetic and kinetic energies defined by
\begin{align*}
&P_B(k) \equiv  \frac{1}{8\pi}\int \tilde{B}_i \tilde{B}^{i*}k^2 d\Omega_k,\nonumber\\
&P_K(k) \equiv \frac{1}{2} \int \tilde{(\sqrt{D}v_i)}\tilde{(\sqrt{D}v^i)}^* k^2 d\Omega_k,
\end{align*}
where the symbols $\tilde{\cdot}$ and $\tilde{\cdot}^*$ denote the Fourier component and its complex conjugate, respectively. $D=\sqrt{\gamma} \rho w$ is the conserved mass density, and $d\Omega_k$ is the solid angle in the $k$-space. The integration of $P_B$ and $P_K$ with $k$ yields the electromagnetic and kinetic energies, respectively. We use simulation data in the nested domains $13$--$16$ and stitch the PSD in each domain to obtain the full PSD spanning three orders of magnitude in $k$ (see the Supplemental Material for details). 

Figure~\ref{fig:PSD} plots $P_B$ and $P_K$ as functions of the post-merger time. Once the two neutron stars come into contact, the shear layer appears at the contact interface, and the Kelvin-Helmholtz instability sets in (see Supplemental Material). The instability is clearly seen in the high wavenumber region of the kinetic PSD, with $k>30$~\footnote{The wavenumber $k$ is normalized by $2\pi/L$ with $L=70$~km throughout this Letter. The physical scale is calculated by $70/k$~km given $k$.}. Around $t-t_\mathrm{merger}\approx 1$--$1.5$~ms, the Kolomogorov spectrum with $\propto k^{-5/3}$ is established. The resultant Kelvin-Helmholtz vorticity amplifies the magnetic field, and the Kazantsev spectrum with $\propto k^{3/2}$ is established in the magnetic PSD by this time. 
It indicates a kinematic dynamo in which the Lorentz force has no back reaction on the flow yet. Due to magnetic-field dissipation at small scales, the peak wavenumber $k_\mathrm{peak}$ of the magnetic PSD remains constant, as expected from the Kazantsev resistive model, until non-linear effects and saturation set in~\cite{Kazantsev:1968,Kulsrud:1992}. The kinetic PSD amplitude in the high wavenumber region with $k\ge 400$ starts to decrease for $t-t_{\rm merger}\gtrsim 1$--$1.5$~ms, indicating that the shear layer starts to be dissipated. Interestingly, the magnetic PSD continues to increase even after this time, and the wavenumber at the peak amplitude $k_{\rm peak}$ shifts towards the larger scale. 

The left panel of Fig.~\ref{fig:PSD_k} shows how $k_{\rm peak}$ evolves with time. Before the merger, it stayed at $\approx 7.2$ corresponding to the stellar scale. 
$k_{\rm peak}$ steeply increases up to $\approx 500$ due to the Kelvin-Helmholtz instability, and decreases to $\approx 150$ after $t-t_{\rm merger} \gtrsim 2$~ms. This evolution can be explained by the selective decay of the turbulent magnetic field~\cite{Schekochihin:2020aqu}. Indeed, when the electromagnetic energy becomes comparable to the kinetic energy of the small-scale eddies, the magnetic field disrupts them, moving the end of Kolmogorov's cascade towards larger scales. 
This leads to a slower growth rate of the magnetic field. Therefore, the growth cannot compensate for dissipation at small scales anymore. However, the larger-scale magnetic field can continue to grow because the dissipation rate, i.e., $\eta k^2$, where $\eta$ represents the resistivity, is lower at larger scales. 
This leads to the shift towards larger scales in the magnetic PSD. We model this behavior by a simple semi-analytical model (see Supplemental Material), and it reasonably agrees with the evolution of $k_{\rm peak}$ in Fig.~\ref{fig:PSD_k}.

The right panel of Fig.~\ref{fig:PSD_k} plots the magnetic PSD with selected wavenumbers $k$ as a function of the post-merger time. The PSD with high wavenumber (the green, orange, and red curves) exponentially grows for $0 \lesssim t-t_{\rm merger} \lesssim 2$~ms. The above picture also explains why the PSD at $k=100$ (green curve) continues to increase at a slower rate for $t-t_\mathrm{merger} \gtrsim 2~{\rm ms}$: the eddies at this scale are not damped by the Lorentz force or dissipated yet and continue to amplify the magnetic field. 
Very interestingly, the PSD at the stellar scale (the cyan curve) starts to exponentially grow at $t-t_{\rm merger} \gtrsim 1$~ms. This delay can be explained by the time that the magnetic field amplified by the kinematic dynamo with the Kazantsev component's scaling of $k^{3/2}$ exceeds the initial magnetic field. It can be seen on the left panel of Fig.~\ref{fig:PSD} where, at $t-t_{\rm merger}\approx 1$~ms (the cyan curve), the Kazantsev component of the spectrum overcomes the initial magnetic field that does not increase before $t-t_\mathrm{merger} \le 1~{\rm ms}$. 
The magnetic PSD at the stellar scale increases by a factor of $\approx 10^5$ until $t-t_{\rm merger}\approx 5$~ms. This increase is explained neither by the compression in which the magnetic field is increased by $O(1)$ because it should be proportional to the power of the averaged-rest mass density, $\bar{\rho}^{2/3}$, nor by the winding of the initial field in which the magnetic-field strength is amplified by $2\pi/(1~{\rm ms}) \times 5~{\rm ms} \approx 30$ where we apply the typical rotational period of $1~{\rm ms}$ for the merger remnant.
It indicates the stellar-scale initial magnetic field of $3.16 \times 10^{12}~{\rm G}$ is amplified up to $1.0 \times 10^{15}~{\rm G}$. Namely, a magnetar is born. 

{\it Discussion}--
The evolution of the magnetic PSD in our simulation is linked to the dissipation at the grid scale. In reality, the dissipation scale, i.e., the microphysical resistivity $\eta$, would be much smaller than the viscous scale, with the viscosity coming from neutrino diffusion $\nu_\nu$ or the microphysical one $\nu$. We are therefore in the regime of a large magnetic Prandtl number ${\rm Pm}=\nu/\eta$. In this regime, if we assume that eddies at the wavenumber $k$ are disrupted when the electromagnetic energy at or above $k$ becomes comparable to the eddy energy, then $k_\mathrm{peak}$ of the magnetic PSD would remain at very small scales with $k\sim 10^{10}$ (see Supplemental Material). However, magnetic reconnection and tearing instabilities in a turbulent flow can change this picture (see, for example, Ref.~\cite{Schekochihin:2020aqu}). Several theoretical mechanisms, such as fast or stochastic reconnection and field-line reconfiguration, could lead to a shift of the magnetic dissipation scale to larger scales through an inverse cascade. In particular, Ref.~\cite{Tripathi:2026} showed the existence of a dynamo by jets from Kelvin-Helmholtz instability that would generate magnetic fields on the scale of a few meters.

According to our models that extrapolate to the astrophysical regime, the magnetic field at the stellar scale could be overpredicted by the simulation presented in this Letter by a factor of $10$, assuming the turbulent resistivity scenario is the case (see Supplemental Material). Nevertheless, the amplification by the Kelvin-Helmholtz would still give a magnetic field of $10^{14} \rm~G$ at the stellar scale, which is enough to affirm that a magnetar is born after only a few milliseconds. However, we would like to point out that this stellar-scale magnetic field is not a dipole and is situated around the equatorial plane. Another mechanism, namely the magnetorotational instability, is required to further amplify into a dipole or a large-scale magnetic field at the poles in order to launch a relativistic jet.

We compare our magnetic PSD with that reported in the global gradient sub-grid scale simulations~\cite{Aguilera-Miret:2025nts}. Irrespective of the binary neutron star models, $k_{\rm peak}$ during the kinematic phase appears at $\approx 110$, which is significantly lower than $k_{\rm peak}$ reported in this Letter. The magnetic PSD amplitude at the stellar scale $k\approx 7.2$ is amplified by $\sim 10^3$ from $t-t_\mathrm{merger}\approx 2.5$~ms to $5$~ms (see their Fig.~4). Their PSD continues to grow until $t-t_{\rm  merger}=10$~ms. This indicates the spectral evolution in the gradient sub-grid scale model is significantly slower than that reported in this Letter. 

Reference~\cite{Gutierrez:2026ngt} reports a zoom-in magnetohydrodynamics simulation of the binary neutron star merger, in which the resolution of $\approx 2.9$~m is achieved in a domain with $L\approx 5.9$~km. They also reported the emergence of the Kolmogorov spectrum in the kinetic PSD and the Kazantsev spectrum in the magnetic PSD at $2.5$~ms after the merger. However, the amplification factor of the electromagnetic energy remains $\approx 10^5$ (see their Fig. 2). On the other hand, it is $\approx 10^9$ in this Letter. 
The most likely reason for the discrepancy may stem from the targeted simulation region, i.e., $L=5.9~{\rm km}$ in Ref.~\cite{Gutierrez:2026ngt} and the entire binary neutron star merger remnant, i.e., $L \approx 70~{\rm km}$ in this Letter. Namely, the back reaction beyond $5.9~{\rm km}$ is not allowed in the former, but allowed in this Letter. 
Furthermore, because of the zoom-in limitation, they cannot access the magnetic PSD at the stellar scale. 

After the Kelvin-Helmholtz instability phase, we expect the magneto-rotational instability to be responsible for generating the turbulence~\cite{Balbus-Hawley:1991}. Due to efficient amplification via the Kelvin-Helmholtz instability, the neutrino viscosity and dragging effect will not reduce the growth rate of the ideal magnetorotational instability~\cite{Guilet:2016sqd,Kiuchi:2023obe}. As reported in Ref.~\cite{Kiuchi:2023obe}, the electromotive force due to the magnetorotational instability-driven turbulence activates the $\alpha\Omega$ dynamo~\cite{Brandenburg:2004jv}, which results in further strong large-scale magnetic field generation. As a matter of fact, we confirm that at the end of the simulation the magnetorotational instability quality factor exceeds the critical value of $10$--$15$~\cite{Hawley:2011tq} inside the relevant region of the merger remnant, and the mean poloidal magnetic field stars to be generated, which is the sign of the $\alpha\Omega$ dynamo. We plan to continue the simulation.

{\it Summary}.--
In this Letter, we report an ab initio global numerical relativity simulation of the binary neutron star merger by employing unprecedented spatial grid resolution of $6.25$\,m. 
Due to the Kelvin-Helmholtz instability, first the small-scale magnetic field exponentially grows, and it drives the subsequent exponential growth in the stellar-scale magnetic field. 
Our result suggests that a magnetar formation could be inevitable in binary neutron star mergers at least temporarily following the mergers, unless it promptly collapses to a black hole~\cite{Metzger:2010pp}. 


\begin{figure*}[t]
 	 \includegraphics[width=0.49\linewidth]{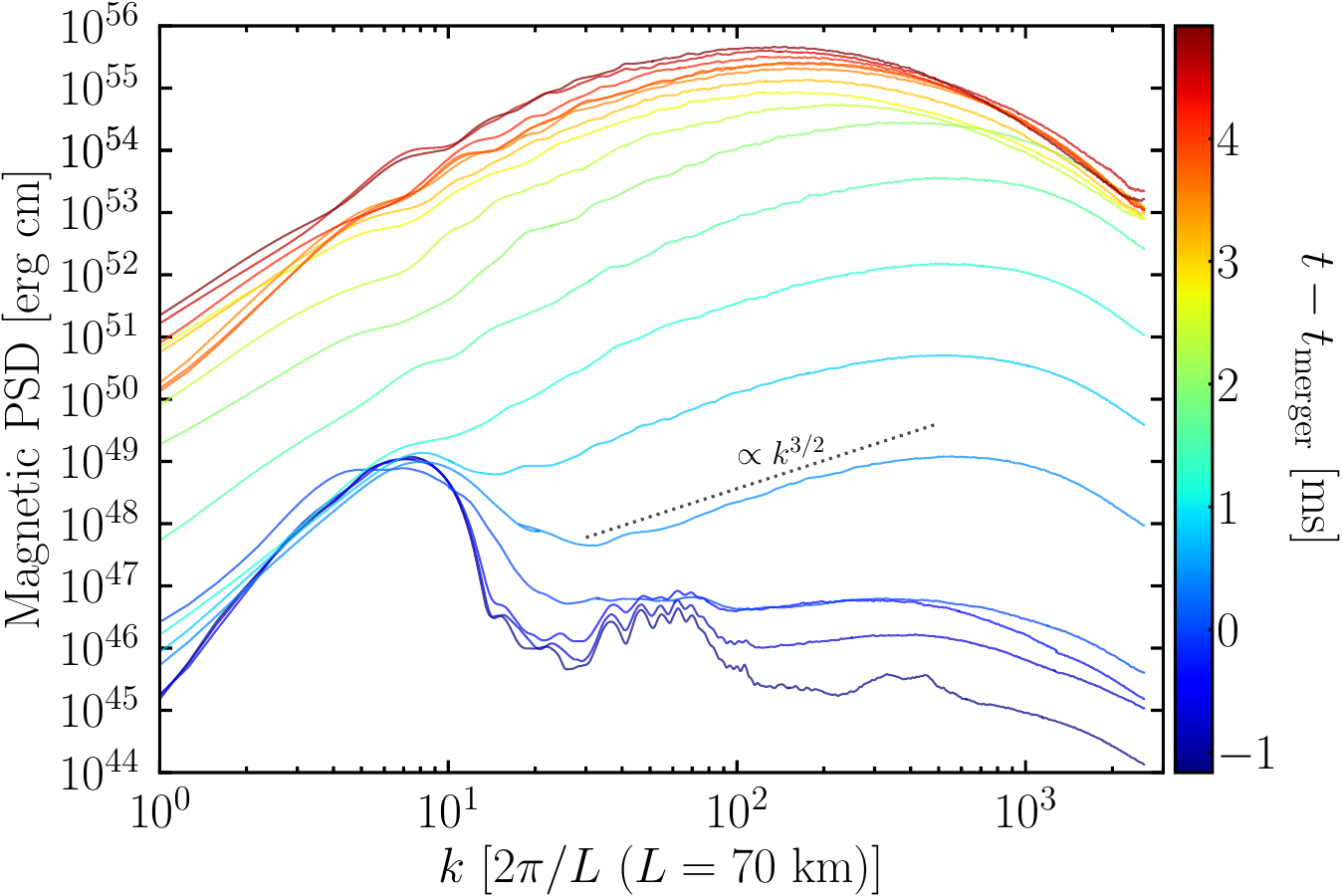}
      	 \includegraphics[width=0.49\linewidth]{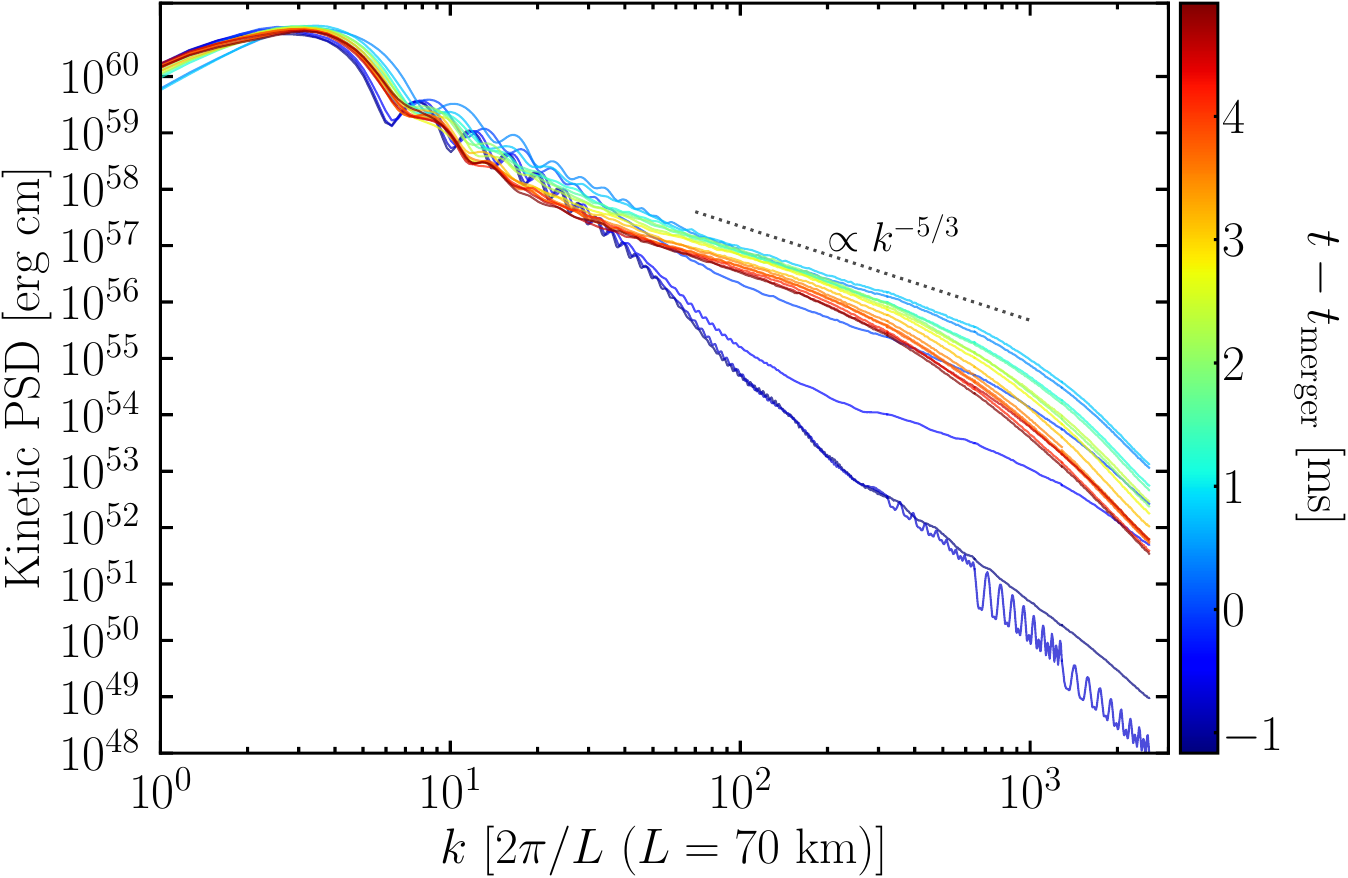}      
 	 \caption{ (Left) Magnetic PSD as a function of the post-merger time. 
     The Kazantsev spectrum proportional to $k^{3/2}$ is shown. (Right) Kinetic PSD as a function of the post-merger time. The Kolmogorov spectrum, proportional to $k^{-5/3}$, is shown.
      }\label{fig:PSD}
\end{figure*}

\begin{figure*}[t]
 	 \includegraphics[width=0.49\linewidth]{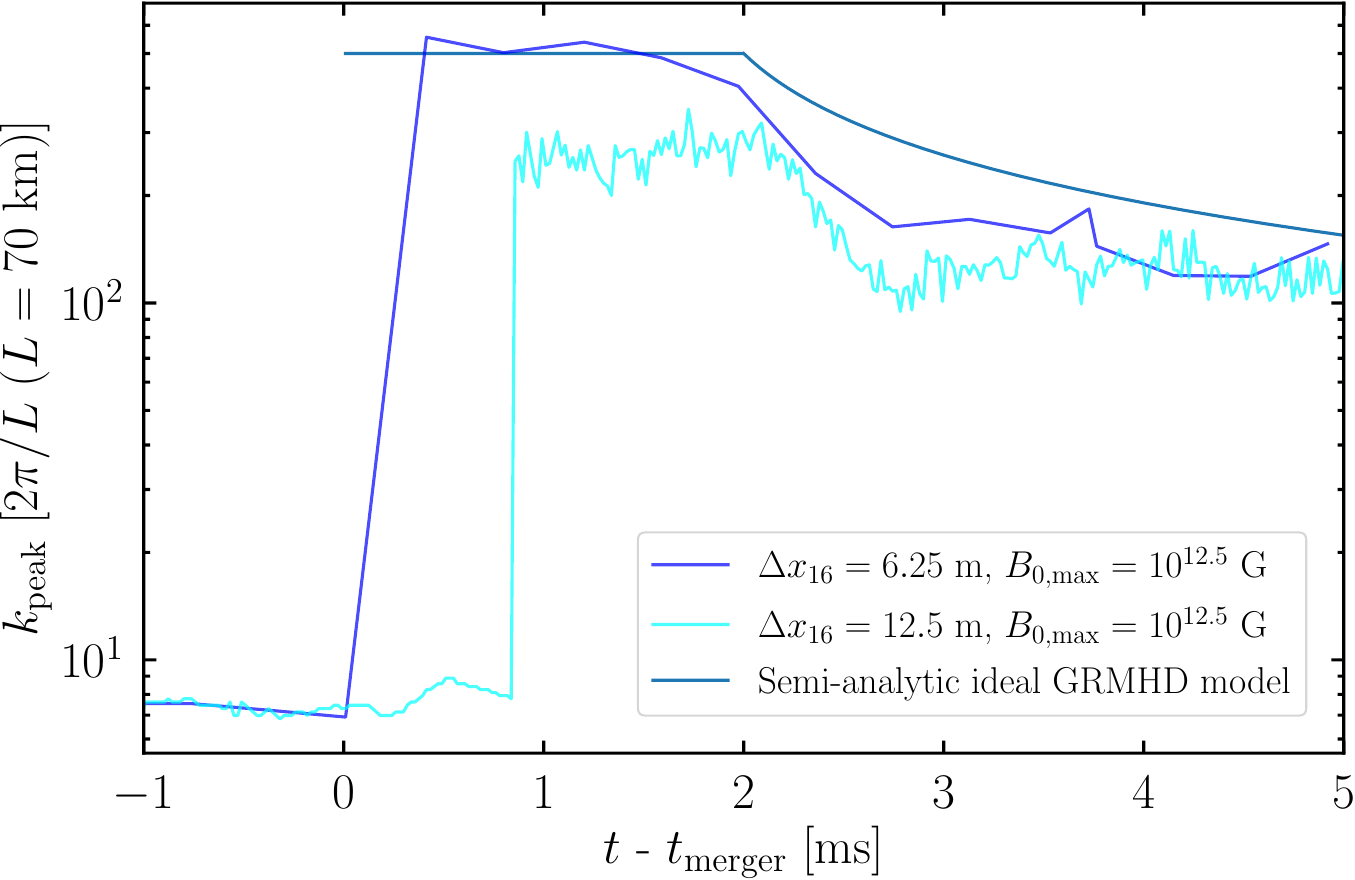}
      	 \includegraphics[width=0.49\linewidth]{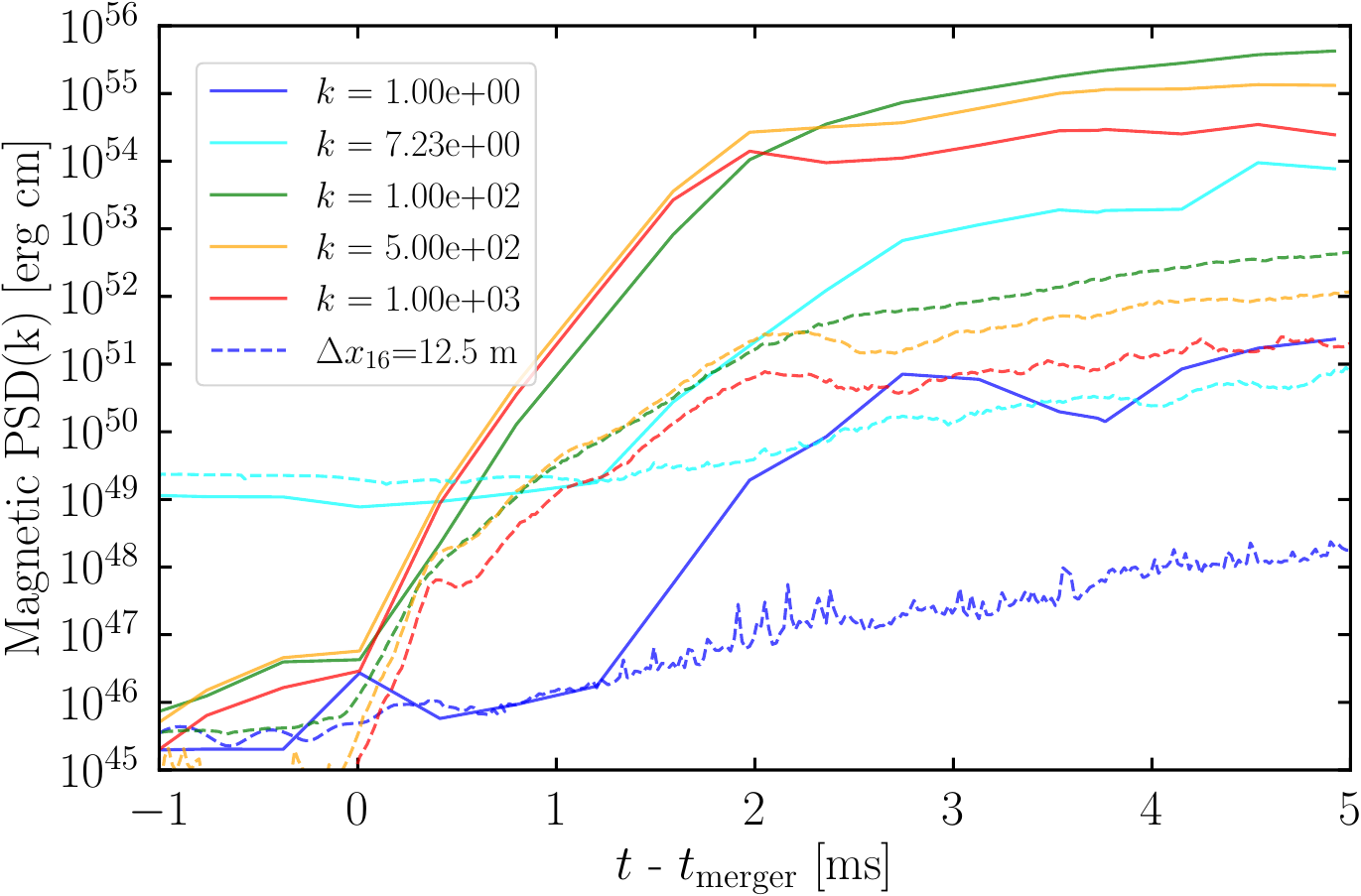}      
 	 \caption{ (Left) Wavenumber at the peak amplitude of the magnetic PSD as a function of the post-merger time. The olive curve represents a semi-analytic model for $k_{\rm peak}$ evolution.  The cyan curve is for the low resolution. 
     (Right) Magnetic PSD with selected wavenumbers as a function of the post-merger time. The dashed curves are for the low resolution.
      }\label{fig:PSD_k}
\end{figure*}

{\it Acknowledgments}.--This work used computational resources of the supercomputer FUGAKU provided by RIKEN through the HPCI System Research Project (Project ID: hp240532, hp250570,hp250066, hp260063). The simulation was also performed on Sakura, Momiji, Raven, and Viper clusters at the Max Planck Computing and Data Facility. This work was in part supported by the Grant-in-Aid for Scientific Research (grant Nos. 23K25869 and 23H04900) of Japan MEXT/JSPS. Kiuchi thanks the Computational Relativistic Astrophysics members, especially Kota Hayashi and Sho Fujibayashi, at AEI for a stimulating discussion. 

\newpage

\appendix
\section{Supplemental Material} 
\subsection{Evolution of the system}
\begin{figure*}[t]
 	 \includegraphics[width=0.99\linewidth]{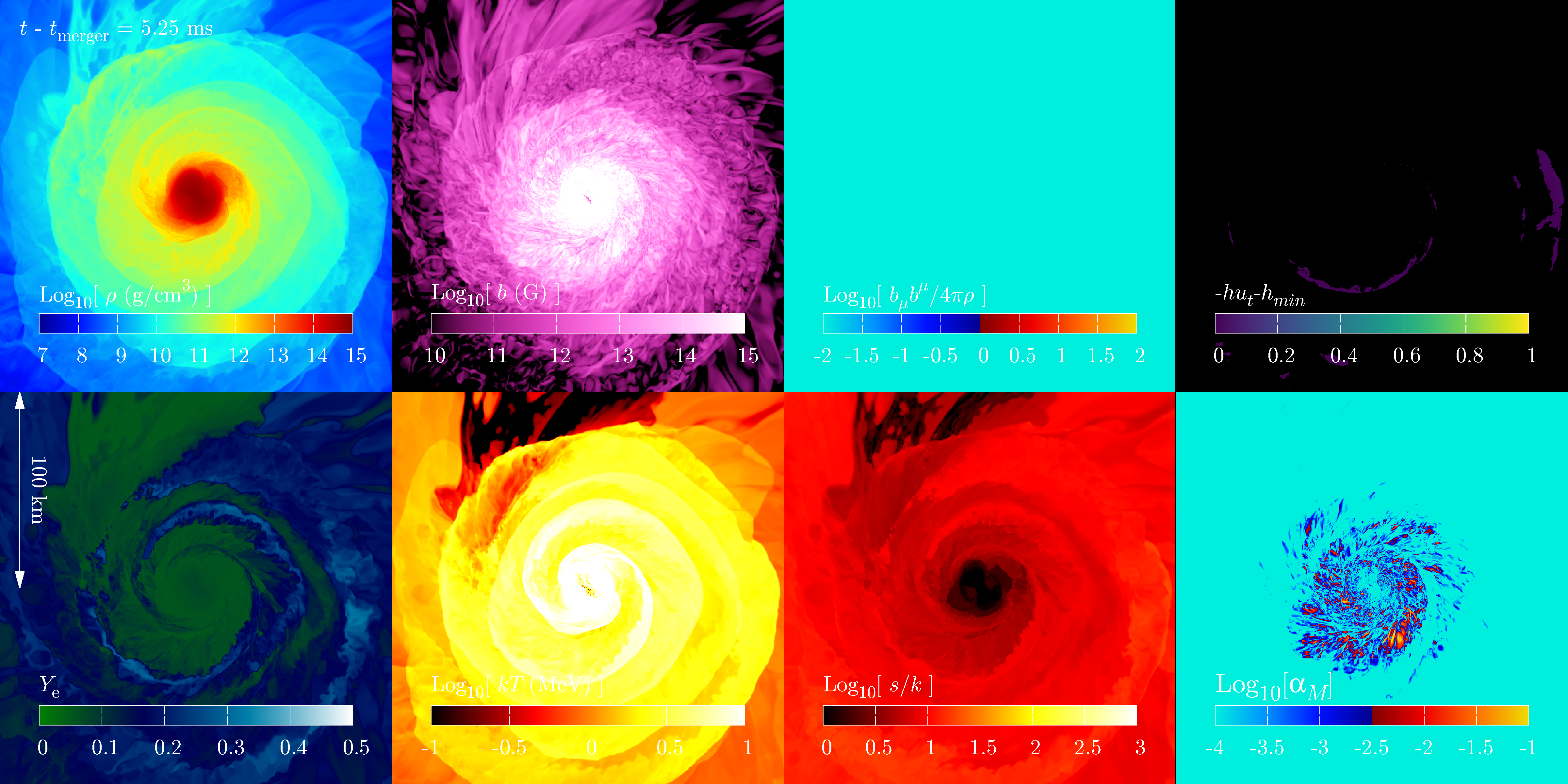} 
     \includegraphics[width=0.99\linewidth]{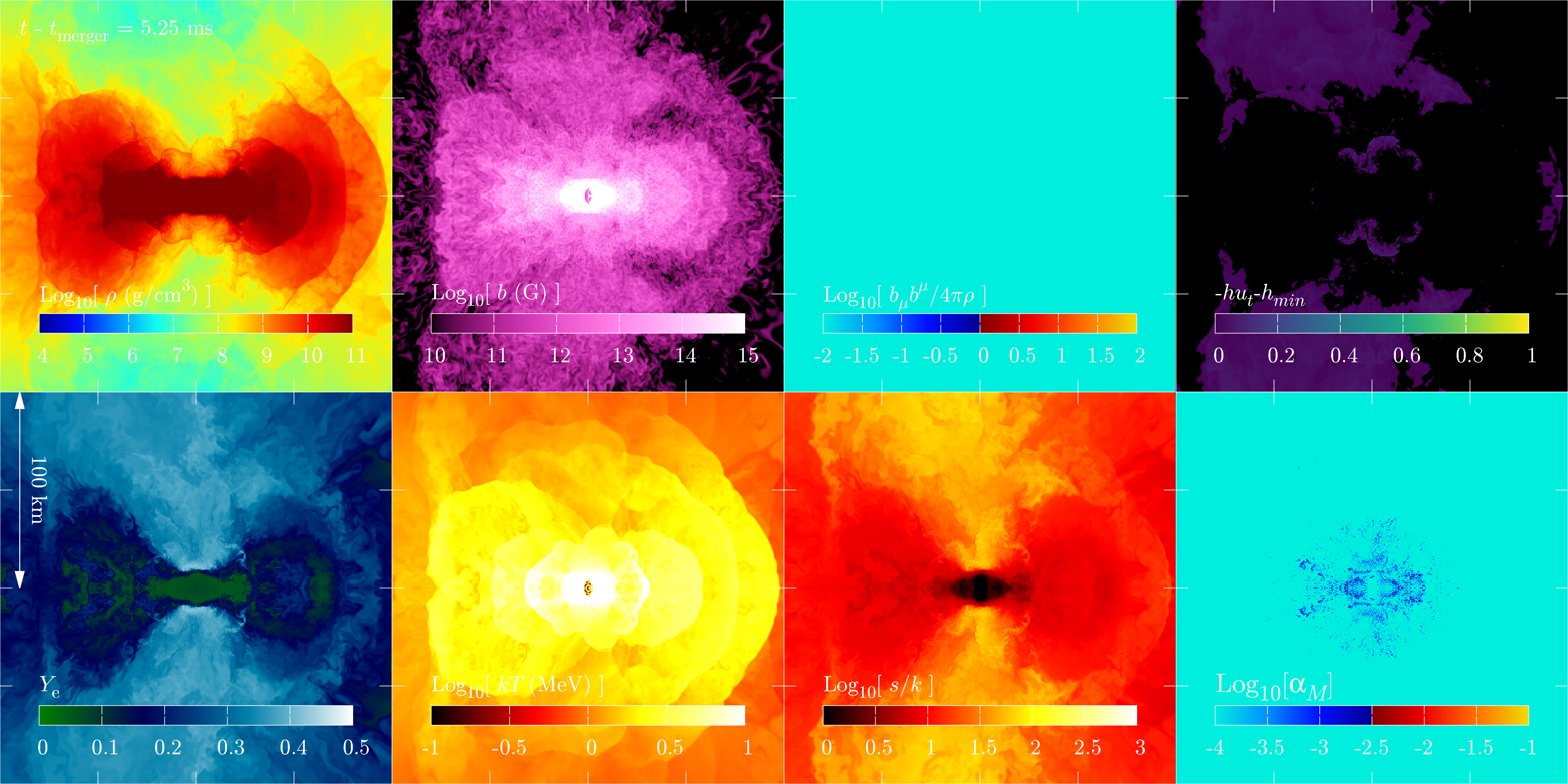} 
 	 \caption{Snapshot of an orbital plane (top) and a meridional plane (bottom) at $t-t_{\rm merger} \approx 5$~ms. From the top-left to the down-right panel, the rest-mass density, the magnetic-field strength, the magnetization parameter, the unboundness of the fluid with the Bernoulli criteria, the electron fraction, the temperature, specific entropy, and the Shakura-Sunyaev parameter, respectively. 
     (see \url{http://www2.yukawa.kyoto-u.ac.jp/~kenta.kiuchi/anime/FUGAKU2026B/out_yuv420p_xy.mp4} and \url{http://www2.yukawa.kyoto-u.ac.jp/~kenta.kiuchi/anime/FUGAKU2026B/out_yuv420p_xz.mp4} for the visualization.) 
         }\label{fig:2D_plot}
\end{figure*}

Figure~\ref{fig:2D_plot} is a two-dimensional slice of several physical quantities. The upper and lower panels are for an orbital plane and a meridional plane, respectively. 
In each panel, from the top-left to the bottom-right corner, the rest-mass density, the magnetic-field strength, the magnetization parameter, the unboundness of the fluid with the Bernoulli criterion, the electron fraction, the temperature, the specific entropy, and the Shakura-Sunyaev parameter are shown, respectively (see also the link for the visualization).

\subsection{Power spectral density stitch}
Since the Fast Fourier Transform (FFT) assumes the equi-distance interval, it is non-trivial to calculate the entire power spectral density (PSD) spanning from the wavenumber corresponding to the largest nested domain size to that corresponding to the Nyquist wavenumber in the finest nested domain. In the literature, to avoid this problem, the interpolation from a coarser domain to a finer domain is employed. However, it may violate energy conservation and, more seriously, affect the slope of the PSD.
 
\begin{figure*}[t]
 	 \includegraphics[width=0.49\linewidth]{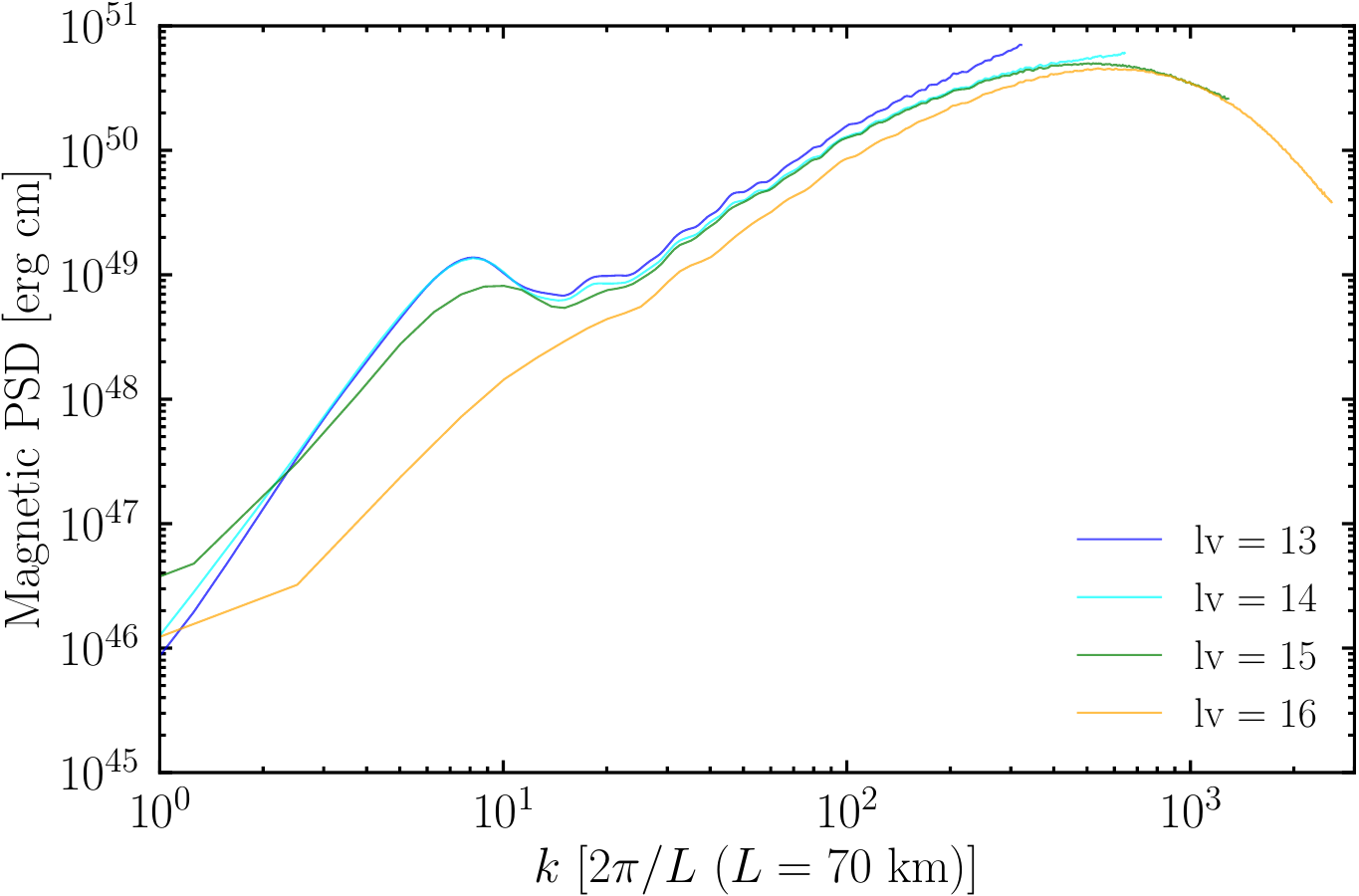}
     \includegraphics[width=0.49\linewidth]{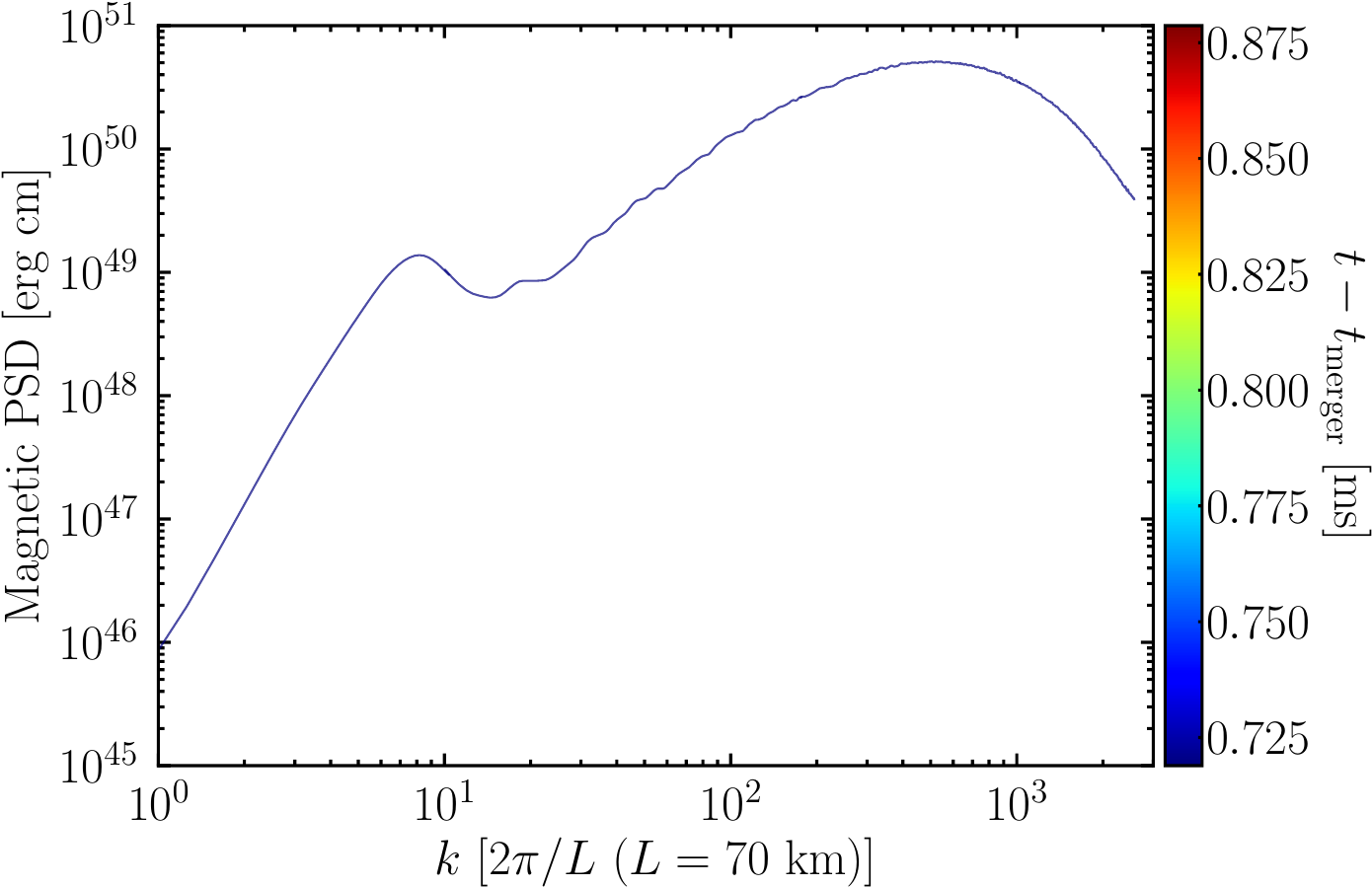}
 	 \caption{(Left) The magnetic PSD in the nested domain ${\rm lv}=13$--$16$. (Right) The stitched magnetic PSD. 
         }\label{fig:PSD_stich}
\end{figure*}

In this paper, we develop a novel scheme to compute the PSD across several nested domains. First, we compute the PSD of the field in each nested domain. We use the Python package {\tt scipy.fft} for the FFT, and determine the normalization such that Parseval's theorem holds. 
Figure~\ref{fig:PSD_stich} plots the PSDs in the nested domains from ${\rm lv}=13$ to $16$ (see the main paper for the nested grid structure). 
We employ the Tukey window function. The PSD at the high and low wavenumbers does not coincide in a coarser and a finer domain, e.g., ${\rm lv}=14$ and $15$, which is the consequence that the PSD at the low wavenumber is affected by the nested domain size. However, at intermediate wavenumbers, the PSD in each domain agrees, e.g., $30 \lesssim k \lesssim 300$ for ${\rm lv}=14$ and $15$. 

Therefore, we "stitch" the PSD in this overlapping region. We search for the wavenumber $k_{\rm stitch}$ that minimizes the difference in the PSD between the coarser and finer nested domains. Then, we combine the PSD in the coarser domain for $k\le k_{\rm stitch}$ with that in the finer domain for $k \ge k_{\rm stitch}$. The right panel of Fig.~\ref{fig:PSD_stich} shows the stitched PSD.

\subsection{Magnetic power spectrum evolution}

\begin{figure*}[t]
 	 \includegraphics[width=0.49\linewidth]{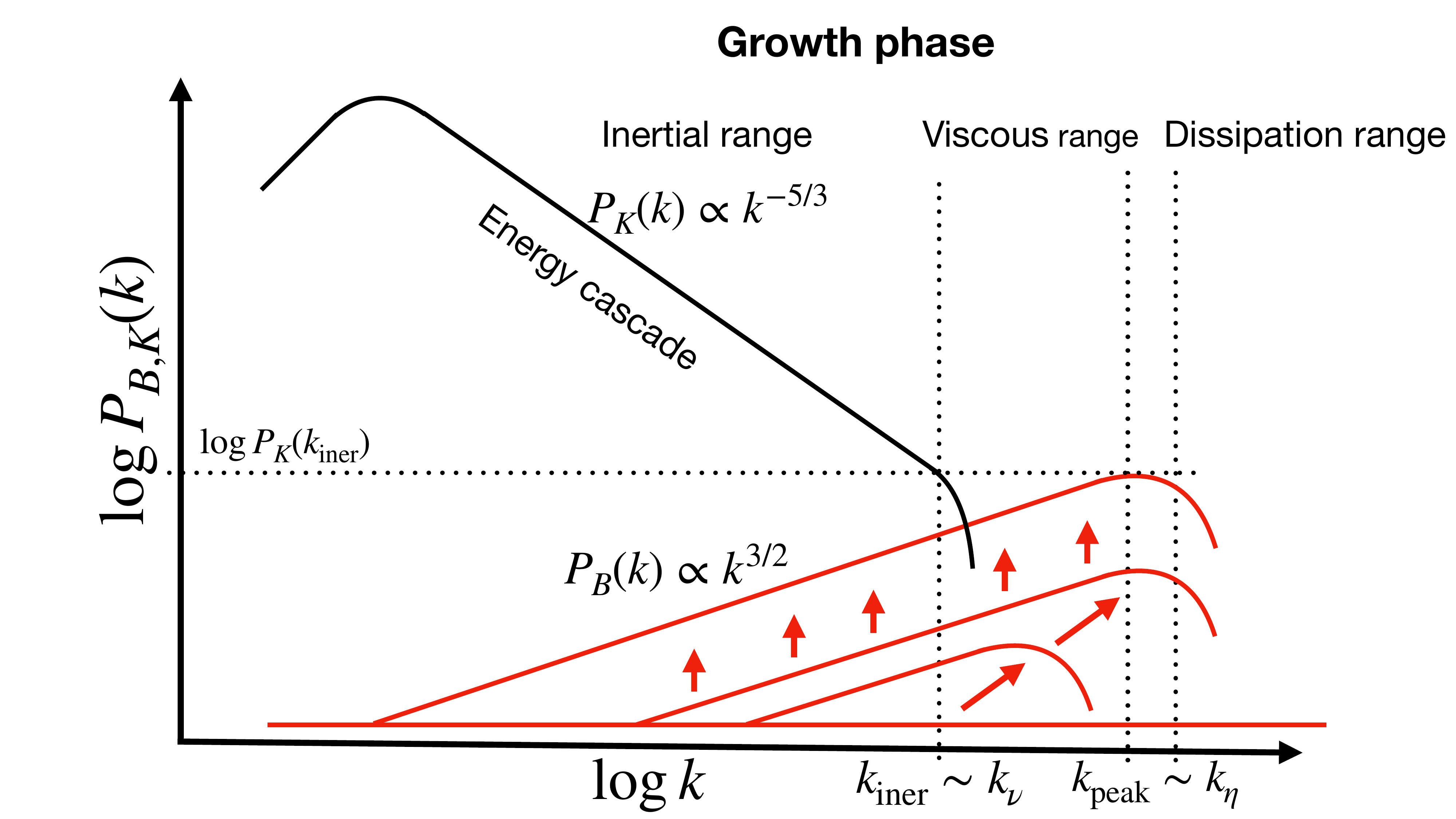} 
     \includegraphics[width=0.49\linewidth]{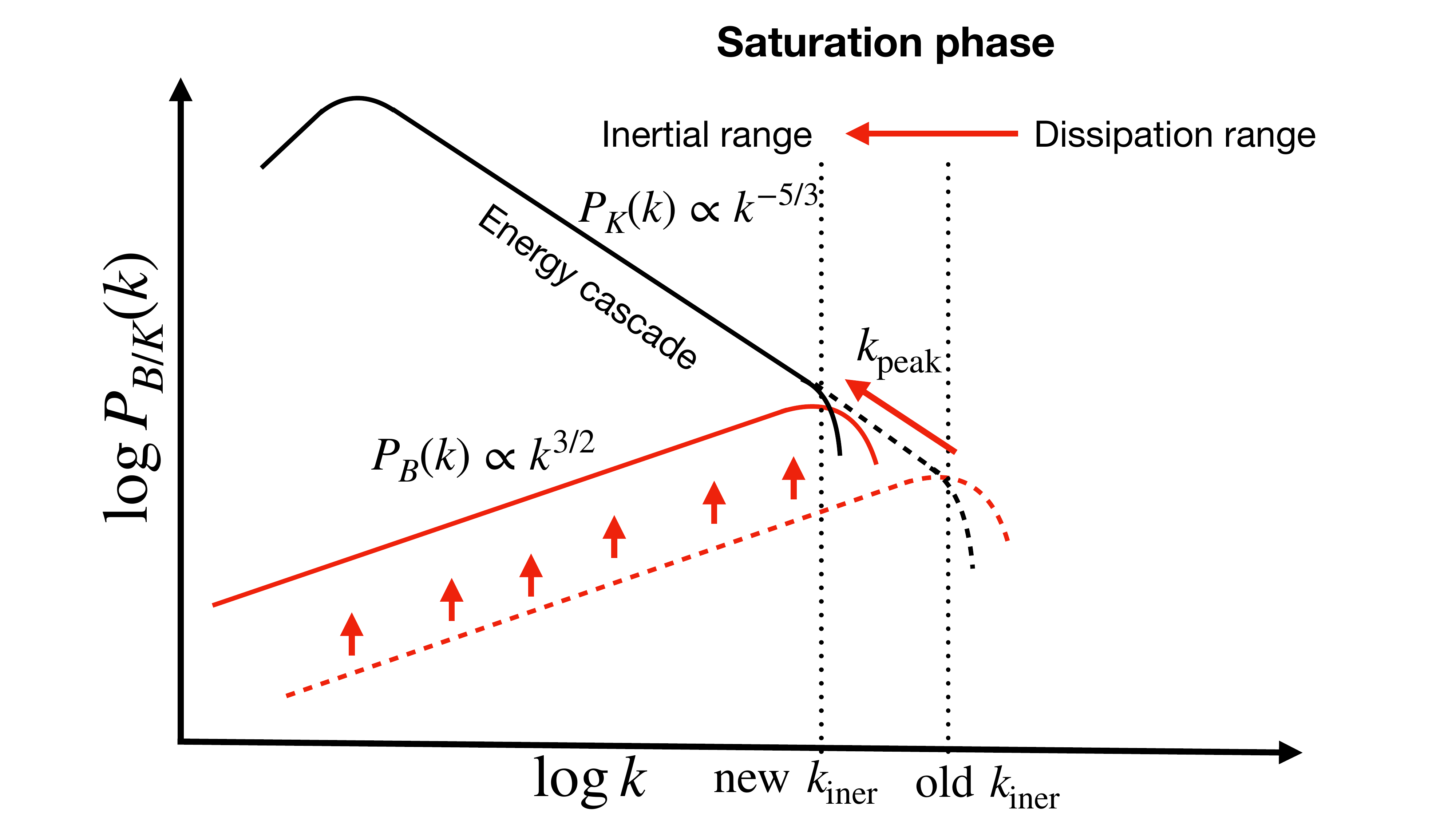} 
 	 \caption{Schematics of the evolution of the magnetic PSD. The left panel is the typical growth of a small-scale dynamo with kinematic growth, with the kinetic PSD in black and the magnetic PSD in red. The right panel is the evolution in the saturation phase, assuming that $k_\mathrm{peak}=k_\mathrm{iner}$. The colors are the same as the previous panel, and the spectra at a previous time are in dashed lines. The red arrow presents the direction of the evolution. 
         }\label{fig:Scheme}
\end{figure*}

In the simulation, we find that the magnetic PSD has characteristics of a kinematic small-scale dynamo, i.e., a magnetic PSD that grows with a scaling of $\propto k^{3/2}$ with a maximum  $k_{\rm peak}$ until the dissipation scale. After $t-t_\mathrm{merger} \approx 2~{\rm ms}$, $k_{\rm peak}$ shifts towards the large scale. To understand this evolution, we explore different dynamo evolution models to saturation to explain the simulation results and extrapolate to the astrophysical regime (See Fig.~\ref{fig:Scheme} for the schematic picture). The kinematic evolution of such a small-scale dynamo with Kolmogorov turbulence is well known~\cite{Schekochihin:2020aqu}.
The magnetic field at given wavenumber $k$ increases with a growth rate $\sigma \propto k_\mathrm{iner}$, where $k_{\mathrm{iner}}$ is the wavenumber where the inertial range of Kolmogorov's turbulence scaling $k^{-5/3}$ ends.
The peak wavenumber $k_\mathrm{peak}$ increases with time, which leads to the scaling $k^{3/2}$ for the magnetic PSD. This continues until $k_\mathrm{peak}$ reaches the dissipation scale and stops growing. The effect of dissipation leads to the growth of the magnetic PSD amplitude with a uniform growth rate $ 2\sigma_\mathrm{peak} \propto k_\mathrm{iner}$ at all scales, which has a peak value at $k_\mathrm{peak}$ (see the left panel of Fig.~\ref{fig:Scheme}). 
The magnetic PSD, therefore, conserves its scaling $k^{3/2}$~\cite{Schekochihin:2020aqu}. 
When the magnetic energy at wavenumbers higher than 
$k_\mathrm{iner}$ becomes similar to the kinetic energy at $k_\mathrm{iner}$, the kinematic regime stops, and the back-reaction of the Lorentz force on the turbulence must be considered. Initially, $k_\mathrm{iner}$ corresponds to the viscous scale $k_{\nu}$. 
This can be written as the following equation:
\begin{widetext}
\begin{align}
    &\int_{k_\mathrm{iner}}^\infty P_B(k,t_\mathrm{sat}) dk = P_\mathrm{kin}(k_\mathrm{iner}), \nonumber\\ 
    &\int_{k_\mathrm{iner}}^{k_\mathrm{peak}} P_{B,0} e^{2\sigma_\mathrm{peak}t_\mathrm{sat}} k^{3/2}dk  +\int_{k_\mathrm{peak}}^\infty P_B(k,t_\mathrm{sat}) dk 
    = P_{K,0} \epsilon_m^{2/3}k_\mathrm{iner}^{-5/3}, \label{eq:gen_k}
 \end{align}
\end{widetext}
where $P_\mathrm{B,0}$ and $\epsilon_m$ are respectively the initial magnetic energy at $t=t_{\rm merger}$ and the Kolmogorov energy injection rate per unit mass that we calibrate on our kinetic PSD. $P_{K,0}$ is the normalization for the kinetic PSD, which has mass divided by the length scale. 
As the magnetic PSD scaling is known only below $k_\mathrm{peak}$ and would decrease rapidly above it, we assume, to simplify, that $\int_{k_\mathrm{peak}}^\infty P_B(k,t_\mathrm{sat}) dk=0$ for all our following models. 

\subsubsection{Ideal GRMHD case}

\begin{figure*}[t]
 	 \includegraphics[width=0.49\linewidth]{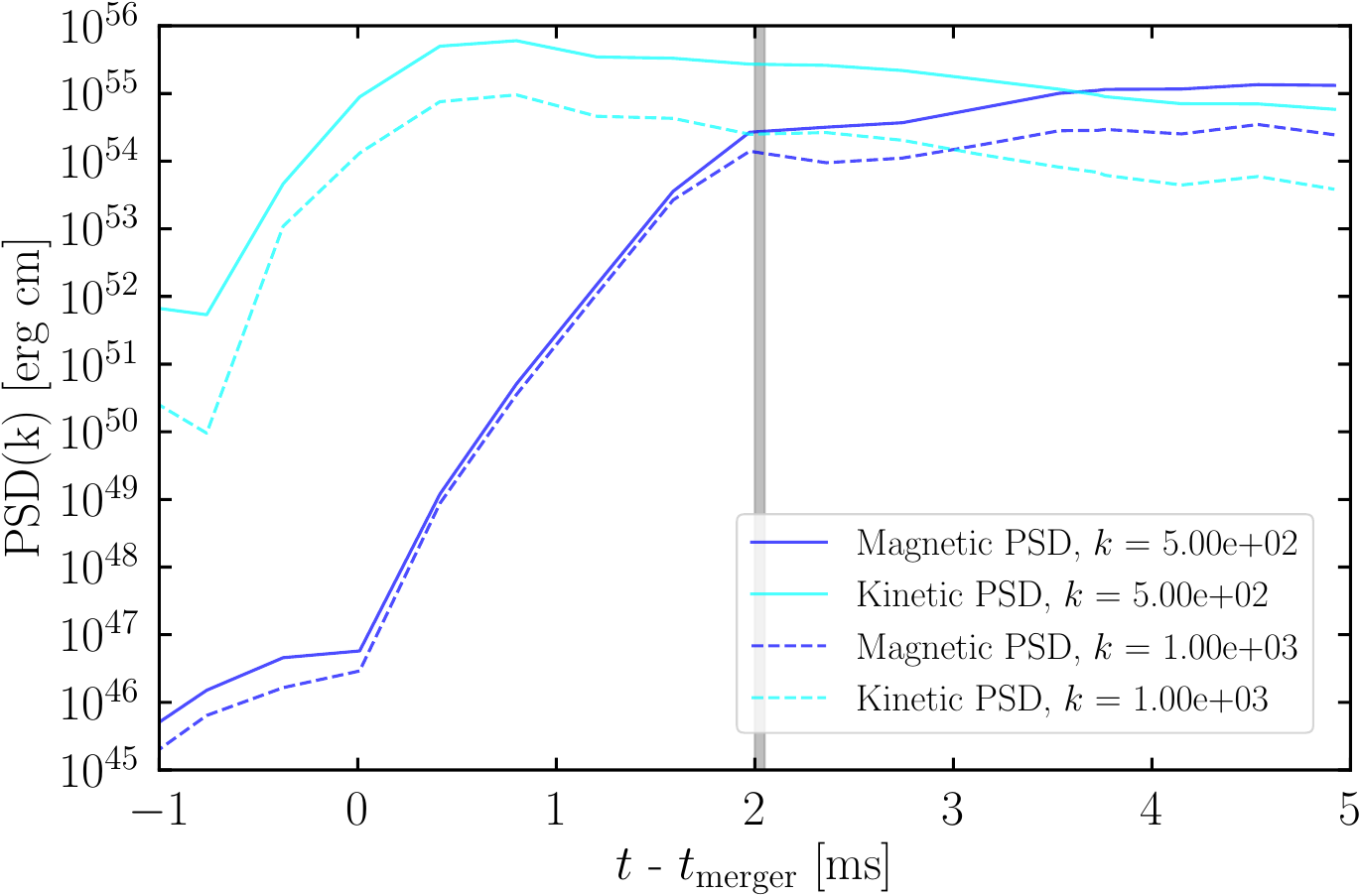}
 	 \caption{Magnetic (blue) and kinetic (cyan) PSD evolution with selected wavenumbers as a function of the post-merger time. 
     The solid and dashed curves are presented for $k=500$ and $1000$, respectively. The vertical gray line presents our estimation of $t_\mathrm{sat}$ (see the text in detail). 
         }\label{fig:k500_1000}
\end{figure*}

We first derive a simple model where the viscous and resistive scales are the same, and we have $\nu=\eta$. This regime is applicable to ideal MHD simulations as the numerical magnetic Prandtl number $P_m=\nu/\eta=O(1)$. As the viscous and resistive dissipation scales are the same, we assume that $k_\mathrm{iner}=k_\mathrm{peak}$. Therefore, the kinematic growth stops when we have equipartition of the magnetic and kinetic energies at $k_\mathrm{iner}=k_\mathrm{peak}$ in Eq.~(\ref{eq:gen_k}):
\begin{align}
    &\tilde{P}_\mathrm{B,0} e^{2\sigma_\mathrm{peak}t_\mathrm{sat}} k_\mathrm{peak}(t_\mathrm{sat})^{3/2}=P_{K,0}\epsilon_m^{2/3}k_\mathrm{iner}(t_\mathrm{sat})^{-5/3}, \label{eq:gen_k2}
\end{align}
where we take the energy at the peak not the integral and introduce $\tilde{P}_\mathrm{B,0}$ to keep the consistency in the dimension.
This equation allows us to estimate $t_\mathrm{sat}$, i.e., when the saturation starts to kick in. Figure~\ref{fig:k500_1000} plots the magnetic and kinetic PSD with $k=500$ and $1000$, respectively~\footnote{The wavenumber $k$ is normalized by $2\pi/L$ with $L=70$~km throughout this supplemental material. The physical scale is calculated by $70/k$~km given $k$.}. It shows that at $t-t_{\rm merger}\approx 2$~ms, after which $k_{\rm peak}$ starts to decrease (see Fig.~3 in the main paper), the nearly equipartition is achieved for $k=1000$. For $k=500$, the sub-equipartition is achieved. Given the flat magnetic PSD with $500 \lesssim k \lesssim 1000$, it justifies Eq.~(\ref{eq:gen_k2}). Also, it gives $t_{\rm sat}\approx t_{\rm merger}+2~{\rm ms}$.

When this equipartition is reached for $k_\mathrm{iner}$ it leads to the disruption of the eddies at this scale, which shifts $k_\mathrm{iner}$ towards lower wavenumbers and larger scales. 
This also means that the growth rate diminishes as $\sigma_\mathrm{peak} \propto k_\mathrm{iner}$, and the magnetic growth at $k_\mathrm{peak}$ cannot be sustained. However, at $k<k_\mathrm{peak}$, the growth can continue as the dissipation rate would be lower. This therefore shifts the peak wavenumber $k_\mathrm{peak}$ of the magnetic PSD and dissipation scales towards larger scales (see the right panel of Fig.~\ref{fig:Scheme}). This would continue until eddies at all scales are disrupted, either by back reaction or by the shear-layer stabilization. 

In order to estimate the time evolution of $k_\mathrm{peak}$, we assume that the equation above must be verified at all times after $t_\mathrm{sat}$ for $k_\mathrm{peak}$, which gives
\begin{align}
    k_\mathrm{peak}(t) = k_\mathrm{peak}(t_\mathrm{sat})e^{-\frac{12}{19} (t-t_\mathrm{sat})Ck_\mathrm{iner}(t)},
\end{align}
where $C$ is a constant for the growth rate $\sigma_\mathrm{peak}=C k_\mathrm{iner}$ that we calibrate on our simulation. The calibration is based on Fig.~3 in the main paper. We measure the growth rate on the right panel with $k_{\rm iner}=k_{\rm peak} =500$ for $0.5~{\rm ms} \lesssim t-t_{\rm merger} \lesssim 1.5~\ \rm ms$. This gives $C\approx 1.0\times 10^{7} {\rm cm~s^{-1}}$. In the low resolution run with $\Delta x_{16}=12.5$~m, $\sigma_\mathrm{peak}=2300~{\rm s^{-1}}$, which gives $C\approx 1.1\times 10^{7} {\rm cm~s^{-1}}$ with $k_\mathrm{peak}=k_\mathrm{iner}=250$ (see Fig.~3 in the main paper).

We therefore have the implicit equation 
\begin{align}
    k_\mathrm{peak}(t) = k_\mathrm{peak}(t_\mathrm{sat})e^{-\frac{12}{19}(t-t_\mathrm{sat}) Ck_\mathrm{peak}(t) },
\end{align}
which can be solved analytically. For that, we define the following variables:
\begin{align}
    &x=k_{\rm peak}(t), a=\frac{12}{19}C (t-t_\mathrm{sat}),b=k_\mathrm{peak}(t_\mathrm{sat}) \nonumber \\
    &\text{ and } u = -ax. 
\end{align} 
After some algebra, the implicit equation is then 
\begin{align}
    -u e^{-u} = a b,
\end{align}
which has the solution 
\begin{align}
    u = -W(a b),
\end{align}
where $W$ is the Lambert W function that verifies $W(z)e^{W(z)}=z$.
This finally gives us 
\begin{align}
 k_{\rm peak}(t) =\frac{ W\left(\frac{12}{19} (t-t_{\rm sat}) C k_{\rm peak}(t_{\rm sat}) \right)}{\frac{12}{19}C (t-t_{\rm sat})} \text{ for } t> t_{\rm sat},
\end{align}
which we plot on Fig.~3 of the main paper where we use $k_{\rm peak}(t)= 500$ for $t\leq t_{\rm sat}\approx t_\mathrm{merger} + 2~{\rm ms}$. In this GRMHD simulation model, $k_\mathrm {peak}(t_\mathrm{sat})$ is proportional to the resolution as shown by Fig.~4 of the main paper and we have $k_\mathrm {peak}(t_\mathrm{sat}) \sim {2\pi}/(10 \Delta x_{16})\times L_{70\rm km}/(2 \pi) \sim 1000 $.

\subsubsection{Uniform magnetic diffusivity case}
In order to extrapolate our simulation results to the astrophysical regime, we now explore the previous scenario in which we consider a uniform magnetic diffusivity much lower than the viscosity (i.e., with a magnetic Prandtl number $P_m\equiv \nu/\eta \gg 1$). 

In the case of a uniform magnetic diffusivity $\eta$, the magnetic dissipation scale would be defined by $\sigma_\mathrm{peak} = C k_\mathrm{iner} = \eta k_\mathrm{\eta}^2$ where we assume the magnetic growth is compensated by the dissipation and therefore $k_\eta= (\frac{C}{\eta} k_\mathrm{iner})^{1/2}$. 
This gives us an upper bound of $k_\mathrm{peak}$ as the magnetic field cannot grow beyond $k>k_\eta$. For $k<k_\eta$, the magnetic field is not dissipated, and it would continue to increase at $k_\mathrm{peak} \sim k_\eta$. Most likely, $k_\mathrm{peak}$ would be smaller than $k_\eta$ by an unknown factor that we assume to be $O(1)$ for simplicity. In this case, the equation for saturation Eq.~(\ref{eq:gen_k}) becomes 
\begin{widetext}
\begin{align}
        P_\mathrm{B,0} e^{2\sigma_\mathrm{peak}t_\mathrm{sat}} \bigg[ \left(\frac{C}{\eta} k_\mathrm{iner}(t_\mathrm{sat})\right)^{5/4} - k_\mathrm{iner}^{5/2}(t_\mathrm{sat})\bigg] =P_{K,0}\epsilon_m^{2/3} k_\mathrm{iner}(t_\mathrm{sat})^{-5/3}, \label{eq:kp2} 
\end{align}
\end{widetext}
As we are in the high magnetic Prandtl number regime, we assume that $k_\eta \gg k_\mathrm{iner}$ (see discussion at the end of this section), so we can neglect the $k_\mathrm{iner}^{5/2}$ term.
After some algebra and assuming that Eq.~(\ref{eq:kp2}) must be verified for $t>t_\mathrm{sat}$, would give 
\begin{align}
    k_\mathrm{iner}(t) = k_\mathrm{iner}(t_\mathrm{sat})e^{-\frac{24}{35} (t-t_\mathrm{sat}) Ck_\mathrm{iner}(t) }. 
\end{align}
This implicit equation can be solved once again with the Lambert W function, and finally gives
\begin{widetext}
\begin{align}
 k_{\rm iner}(t)&= \frac{ W\left(\frac{24}{35} (t-t_\mathrm{sat}) C k_{\rm iner}(t_\mathrm{sat}) \right)}{\frac{24}{35}C (t-t_\mathrm{sat})},\\
 k_{\rm peak}(t) &= \sqrt{\frac{C}{\eta} k_\mathrm{iner}(t)} =\left(\frac{ W\left(\frac{24}{35} (t-t_\mathrm{sat}) C k_{\rm iner}(t_\mathrm{sat}) \right)}{\frac{24}{35}\eta (t-t_\mathrm{sat})}\right)^{1/2} \text{ for } t> t_{\rm sat}.
\end{align}
\end{widetext}
The initial value of $k_\mathrm{iner}(t_\mathrm{sat})$ can either be estimated from the kinetic PSD in the numerical simulation or with neutrino/microphysical viscosity $\nu$. In the latter case and assuming classical Kolmogorov turbulence, we have $k_\mathrm{iner}(t_\mathrm{sat}) \sim  k_{\nu} =2\pi \epsilon_m^{1/4} \nu^{-3/4}$. 
Note that if we assume that the magnetic growth rate corresponds to the inverse of Kolmogorov time scale at $k_\mathrm{iner}$, then the compensation of magnetic growth and dissipation now gives $(\epsilon_m/\nu)^{1/2} = \eta k_{\eta}^{2}$. We then find $k_{\eta}=\sqrt{P_m} k_{\nu}$, which justify our assumption that $k_{\eta}\gg k_\mathrm{iner}$. 
In general, the magnetic field growth rate can be different from the Kolmogorov time scale or the kinetic growth rate of Kelvin-Helmholtz instability, but from local simulations in Ref.~\cite{Tripathi:2026} it would be $O(10)$ times slower, which would not change that $k_{\eta}\gg k_\mathrm{iner}$ when $P_m \gg 1$.

We calibrate $\epsilon_m\approx 8\times 10^{21}~{\rm cm^2~s^{-3}}$ in the kinetic PSD at $k_{\rm peak}=500$ at the saturation time $t_{\rm sat}$ where we assume the remnant massive neutron star mass $\approx 2.7M_\odot$ as the mass scale (see Eq.~(\ref{eq:gen_k})). It nicely agrees with a naive estimate: $\epsilon_m\sim v_0^3/l_0\sim 10^{22}~{\rm cm^2 s^{-3}}$ where $v_0$ and $l_0$ are the velocity and length scale at the energy injection scale in the Kolomogorov spectrum. We set $v_0=0.1{\rm c}$ and $l_0=10^6~{\rm cm}$, respectively.

\subsubsection{Turbulent magnetic diffusivity}

We derive a model for microphysical resistivity in the previous section, but turbulent dissipation or stochastic reconnection can lead to a higher effective resistivity.
In this model, we would like to include turbulent resistivity that may change the dissipation case.
To keep as much generality as possible in the turbulent dissipation/reconnection mechanisms, we use the model developed in Ref.~\cite{Subramanian:1999mr} 
and parametrize the non-linear back reaction on the flow as an increase in the effective resistivity at wavenumbers $k>k_\mathrm{iner}$ 
\begin{align}
    \eta_\mathrm{eff} = \eta + \tau \left\langle \frac{B^2}{4\pi \rho}\right \rangle,
\end{align}
where $\tau$ is an adjustable constant as a time, and $\langle \cdot\rangle$ is a space average over the scale at the end of the inertial range $l_\mathrm{iner}={2\pi}/{k_\mathrm{iner}}$. 
As a first assumption, we assume equipartition between the magnetic field and the turbulent velocity at the scale $k_\mathrm{iner}$: $\left\langle {B^2}/{4\pi \rho}\right\rangle_{k_\mathrm{iner}} = v_\mathrm{iner}^2$. 
We also set $\tau = l_\mathrm{iner}v_\mathrm{iner}^{-1} =\epsilon_m^{-1/3}(k_\mathrm{iner}/(2\pi))^{-2/3}$, which is the turnover time of Kolmogorov turbulence and can be shorter than the magnetic growth time scale. 

However, the magnetic growth can not be faster than the eddy turnover time of Kolmogorov turbulence, as the turnover time is also the stretching rate of magnetic field lines, which would be the case with our scaling $\sigma_\mathrm{peak}\propto k_\mathrm{iner}$ as $1/\tau \propto k_\mathrm{iner}^{2/3}$. 
To avoid this issue at small scales, we set $\sigma_\mathrm{peak}= 1/(C_2^2 \tau)$, where the factor $C_2$ comes from the calibration of the growth rate to match our simulation. By using the values of our simulation $\sigma_\mathrm{peak}=4490~\rm s^{-1}$ with $k_\mathrm{iner}= 500$ and $\tau =\epsilon_m^{-1/3}(k_\mathrm{iner}/(2\pi))^{-2/3}$, we obtain $C_2^2\approx7.67$. We then equate the magnetic growth rate to magnetic dissipation 
\begin{align}
       \sigma_\mathrm{peak} =  \eta_\mathrm{eff} k_{\eta}^2 = \left(\eta + \tau v_\mathrm{iner}^2\right) k_{\eta}^2  
\end{align}
The regime where $\eta \gg \tau v_\mathrm{iner}^2 $ is the one from the previous subsection, so we focus on the opposite regime where $\eta \ll \tau v_\mathrm{iner}^2 $, i.e., when turbulent resistivity dominates over the micro-physical one.
We then obtain that 
\begin{align}
    k_{\eta_\mathrm{eff}} &= \frac{1}{C_2 v_\mathrm{iner} \tau} 
    = \frac{k_\mathrm{iner}}{2\pi C_2},
\end{align}
where $v_\mathrm{iner}= (2\pi)^{1/3} \epsilon_{\rm m}^{1/3}  k_\mathrm{iner}^{-1/3} = v_0 (l_\mathrm{iner}/l_0)^{1/3}$. 
This regime is close to the ideal GRMHD regime that we have explored in the first section. 

The main difference is that now the growth rate scales as $\sigma_\mathrm{peak} \propto k^{2/3}$ and we have $k_{\eta_\mathrm{eff}}<k_\mathrm{iner}$. As $k_{\eta_\mathrm{eff}}<k_\mathrm{iner}$, the growth rate would actually be $\sigma_\mathrm{peak} \propto k_{\eta_\mathrm{eff}}^{2/3}$ as the amplification is at large scale than $k_\mathrm{iner}$. In a similar fashion to the ideal GRMHD regime, we also have to take the energy at the peak and not the integral.
The equation for saturation can then be written in terms of $k_{\eta_\mathrm{eff}}=k_\mathrm{peak}$, and it becomes 
\begin{widetext}
\begin{align}
                  \tilde{P}_\mathrm{B,0} e^{2\sigma_\mathrm{peak}t_\mathrm{sat}} \left(k_\mathrm{peak}(t_\mathrm{sat})\right)^{3/2} 
        =P_{K,0}\epsilon_m^{2/3}\left(2\pi C_2 k_\mathrm{peak}(t_\mathrm{sat})\right)^{-5/3}, \label{eq:kp3}
\end{align}
\end{widetext}
which, after some algebra, would give
\begin{align}
 k_\mathrm{peak}(t) = k_\mathrm{peak}(t_\mathrm{sat})e^{-\frac{12}{19C_2^2} (t-t_\mathrm{sat}) (\epsilon_m/(4\pi^2))^{1/3} k_\mathrm{peak}^{2/3}(t) },
\end{align}
where we assume $\sigma_\mathrm{peak} \propto k_\mathrm{peak}^{2/3}$ instead of $\propto k_\mathrm{iner}^{2/3}$.

To solve the implicit equation, we need to set $k_\mathrm{new}=k_\mathrm{peak}^{2/3}$, which gives
\begin{align}
 k_{\rm new}(t)&= k_\mathrm{new}(t_\mathrm{sat})e^{-\frac{24}{57C_2^2} (t-t_\mathrm{sat}) (\epsilon_m/(4\pi^2))^{1/3} k_\mathrm{new}(t) }.
\end{align}
We then use similar techniques as before to solve the implicit equation in $k_{\rm new}$ and obtain
\begin{widetext}
\begin{align}
     k_{\rm new}(t) &= \frac{ W\left(\frac{24}{57C_2^2} (t-t_\mathrm{sat}) (\epsilon_m/(4\pi^2))^{1/3} k_{\rm new}(t_\mathrm{sat}) \right)}{\frac{24}{57C_2^2}(\epsilon_m/(4\pi^2))^{1/3} (t-t_\mathrm{sat})},\\
    k_{\rm peak}(t) &= k_\mathrm{new}^{3/2}(t) 
    = \left(\frac{ W\left(\frac{24}{57C_2^2} (t-t_\mathrm{sat})  (\epsilon_m/(4\pi^2))^{1/3} k_{\rm peak}^{2/3}(t_\mathrm{sat}) \right)}{\frac{24}{57C_2^2}(\epsilon_m/(4\pi^2))^{1/3} (t-t_\mathrm{sat})}\right)^{3/2}\text{ for } t> t_{\rm sat}.
\end{align}
\end{widetext}

Note that in this model, we have neglected the impact of non-linear effects on the growth in order to have the time evolution to see whether the growth is faster than the shear layer duration. We will compare our final results with the model in Ref.~\cite{Schober:2015}, which takes into account non-linear growth but not the duration of the Kelvin-Helmholtz instability. They find the following results at saturation for incompressible Kolmogorov turbulence 
\begin{align}
    k_\mathrm{peak} \approx 100 k_{l_0} \approx 700 \\
    \rho_{B,\mathrm{sat}} \approx 0.15 \rho v_0^2 
\end{align}
where $k_{l_0}$ is the  wavenumber corresponding to the scale $l_0$ and $\rho_{B,\mathrm{sat}}$ is the magnetic energy density at saturation.
The magnetic PSD at the end of the dynamo is therefore
\begin{align}
    P_B(k) = \alpha_\mathrm{nl,sat}k^{3/2}K_0(k/k_\mathrm{peak}),
\end{align}
where $K_0$ is the modified Bessel function of the second kind and $\alpha_\mathrm{nl,sat}$ is defined by the saturation energy through %
\begin{align}
    \rho_{B,\mathrm{sat}} = \int P_B(k) dk.
\end{align}
We will compare these results with our previous models for binary neutron star mergers.

\subsubsection{Application to the astrophysical case}

We now apply the previous models to the astrophysical scenario of a binary neutron star merger. For the dissipation processes, we use the viscosity $\nu_\mathrm{micro}$ of electrons in a neutron star, the viscosity $\nu_\mathrm{neutrino}$ from momentum diffusion by neutrinos in an optically thick regime, and the resistivity due to electron-electron scattering $\eta_e$. The values we use are $\nu_\mathrm{micro}=0.4~\rm cm^2 \ s^{-1}$, $\eta=3.33\times 10^{-4}~\rm cm^2 \ s^{-1}$ from \cite{Thompson:1993} and $\nu_\mathrm{neutrino}=5\times10^{7}~\rm cm^2 \ s^{-1}$ from \cite{Guilet:2016sqd}. 
We then use the estimation of $k_\mathrm{iner}(t_\mathrm{sat})=k_{\nu_i}=2\pi \epsilon_m^{1/4} \nu_i^{-3/4}$ from classical Kolmogorov turbulence, where $\nu_i$ is either from the electrons or the neutrinos. In addition, we also test for the ideal GRMHD model the case where we hypothetically resolve all the scales until resistive dissipation scales, which are given by $k_\mathrm{iner}= k_{\eta} = 2\pi \epsilon_m^{1/4} \eta^{-3/4}$.

\begin{figure*}[t]
 	 \includegraphics[width=0.49\linewidth]{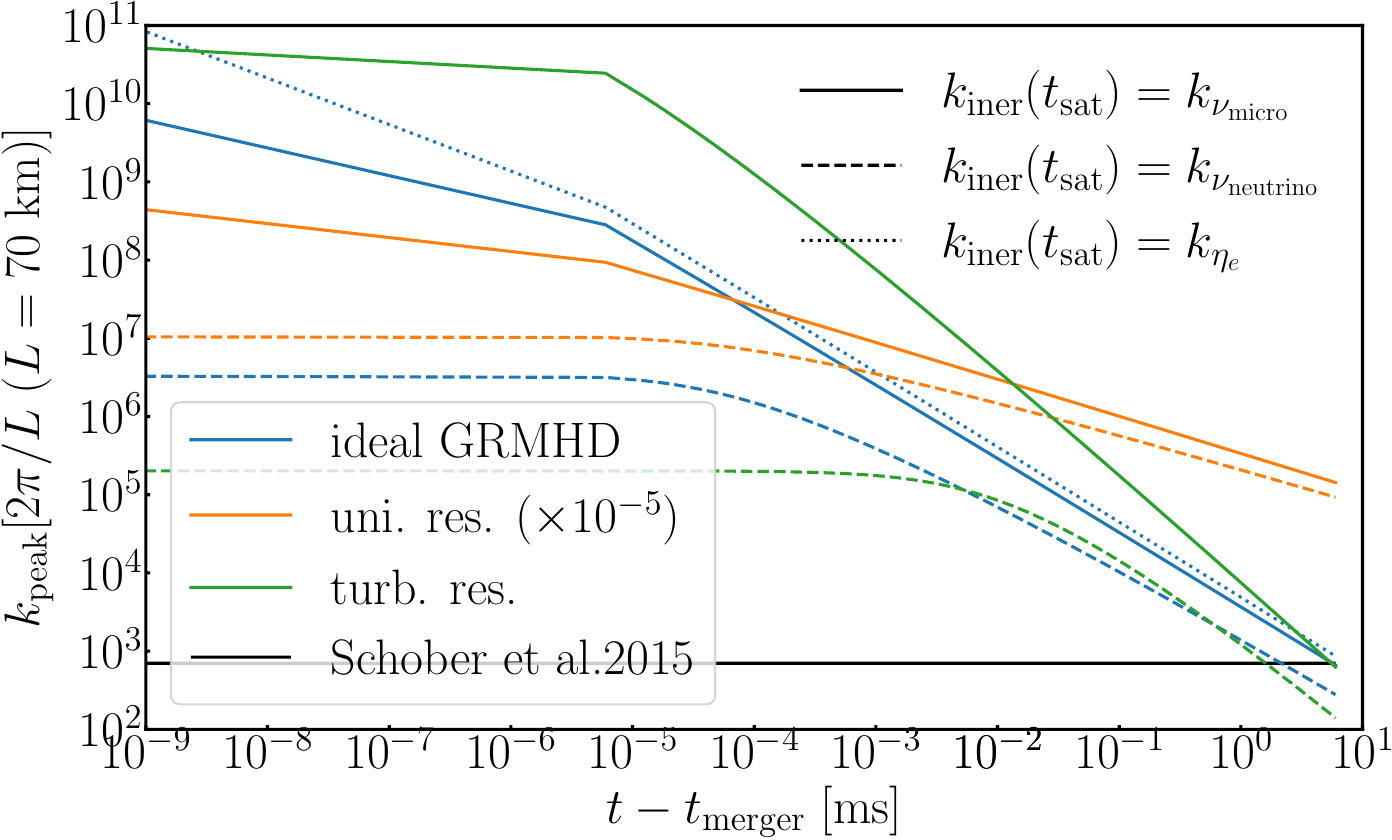} 
     \includegraphics[width=0.49\linewidth]{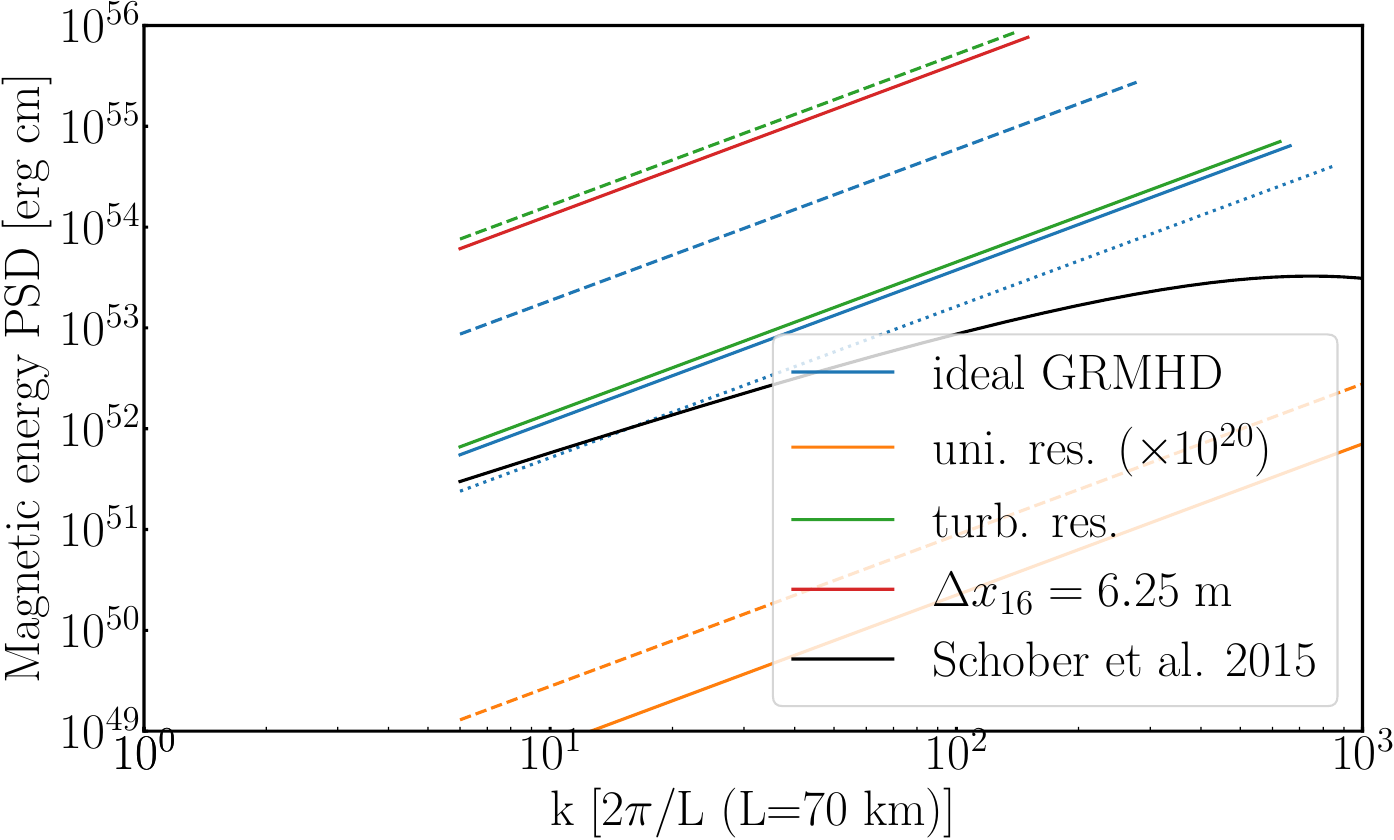}
 	 \caption{(Left) Evolution of $k_\mathrm{peak}$ according to the different models in color with different initial inertial scale of Kolmogorov turbulence due to microphysical viscosity (solid lines), neutrino viscosity (dashed lines). The uniform resistivity case is downsized by a factor of $10^{-5}$. 
     (Right) Predicted magnetic PSD at saturation for the different evolution models of $k_\mathrm{peak}$. The uniform resistivity case is magnified by a factor of $10^{20}$.
         }\label{fig:k_peak_evolution}
\end{figure*}

The left panel of Fig.~\ref{fig:k_peak_evolution} shows the evolution of $k_\mathrm{peak}$ for different models. As expected, the models with uniform resistivity stay at small scales for the whole duration of the Kelvin-Helmholtz instability (the orange curves). However, the turbulent resistivity (the green curves) or ideal GRMHD (the blue curves) evolves from very small scales to a $k_\mathrm{peak}\approx 500-1000$, depending on $k_\mathrm{iner}(t_\mathrm{sat})$, which are given by the hypothetical resolution for the ideal GRMHD case or by the value of viscosity for the turbulent resistive models. The models starting from microphysical viscosity seem to agree nicely at $t-t_\mathrm{merger}=6 \rm~ms$ with the predicted $k_\mathrm{peak}$ in Ref.~\cite{Schober:2015} (the black line).
It is quite interesting to note that $k_\mathrm{peak}$ from ideal GRMHD simulations is not expected to converge, even with a hypothetical resolution higher by a factor $\geq 10^{3}$ (see the solid vs dashed or dotted blue curves) as the resolution is proportional to $k_\mathrm{iner}(t_\mathrm{sat})$. This might be due to the fact that we did not put any lower limit on $k_\mathrm{peak}$ where there would be no dynamo anymore due to a growth rate being too small (or equivalently, a magnetic Reynolds number smaller than the critical magnetic Reynolds number).

Overall, the extrapolation predicts a $k_\mathrm{peak}$ that would be $\sim 5$ times higher than the result of the simulation after the saturation. This difference may be due to a different saturation mechanism from the equipartition of the magnetic field. Indeed, in the simulation, the evolution of the peak might be due to the dissipation of the turbulent eddies when the system becomes stable to the Kelvin-Helmholtz instability. This would explain why the evolution of the peak wavenumbers for two resolutions $\Delta x_{16} = 6.25 \rm m$  and $\Delta x_{16} = 12.5 \rm m$ is similar, while the latter does not reach equipartition between magnetic energy and kinetic energy (see the next subsection).

In the right panel of Fig.~\ref{fig:k_peak_evolution}, we also plot the saturated spectra at $t-t_\mathrm{merger}=6\rm~ms$ using $k_\mathrm{peak}$ found from the previous models and assuming equipartition at $k_\mathrm{peak}$ for the normalization of the magnetic field. It shows that, in the turbulent resistivity case with the microphysical viscosity, the magnetic field energy at the stellar size ($k\sim 7$) is overpredicted by two orders of magnitude compared to the realistic astrophysical situation, which leads to an overestimate of the magnetic-field strength by a factor $10$. 
Note that our time evolution is probably slower than in reality, as we may be underestimating $C$ and $C_2$. Then the saturated spectrum found in Ref.~\cite{Schober:2015} may be reached faster than $\sim 6 \rm~ms$. 



\begin{figure*}[t]
 	 \includegraphics[width=0.49\linewidth]{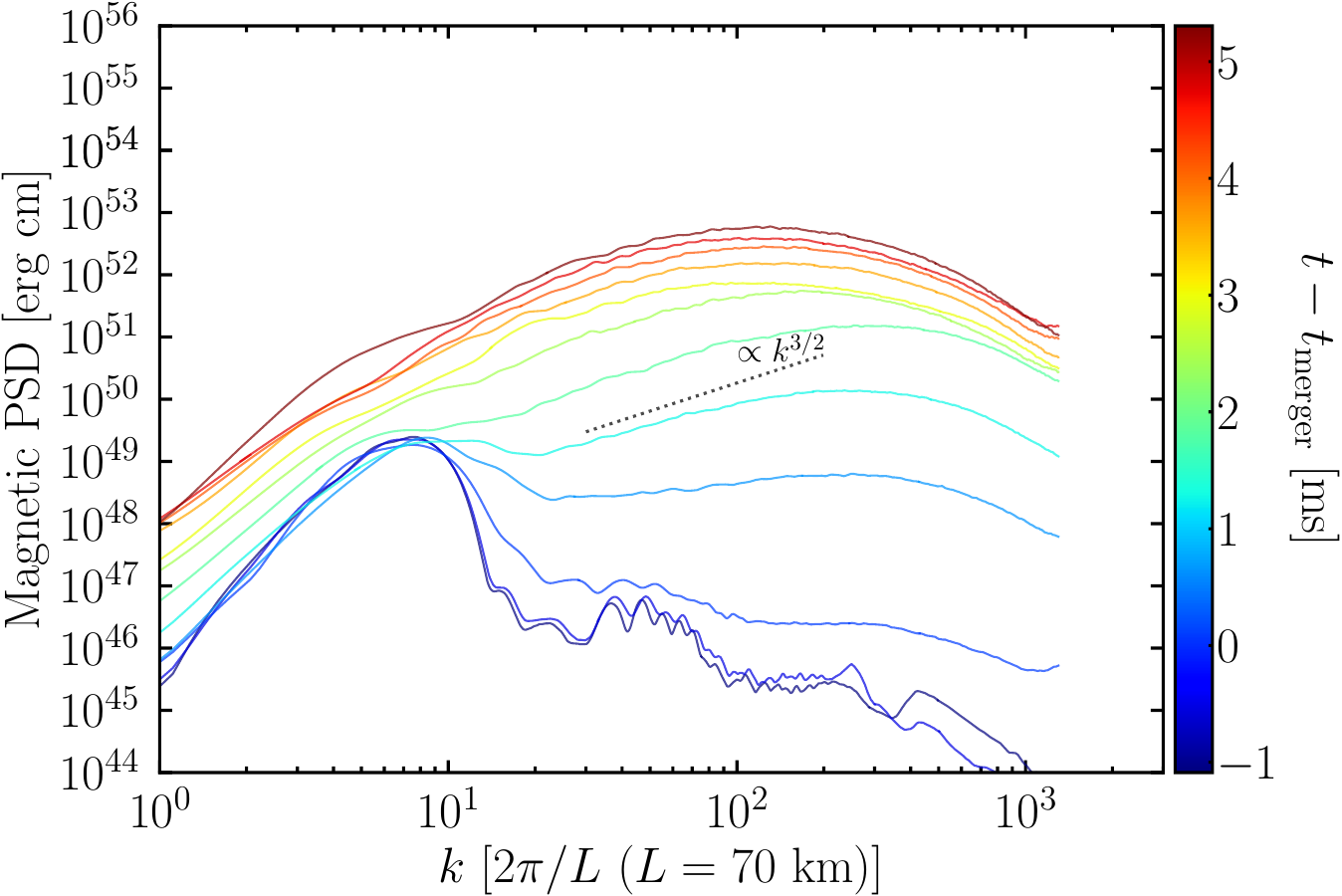}
      	 \includegraphics[width=0.49\linewidth]{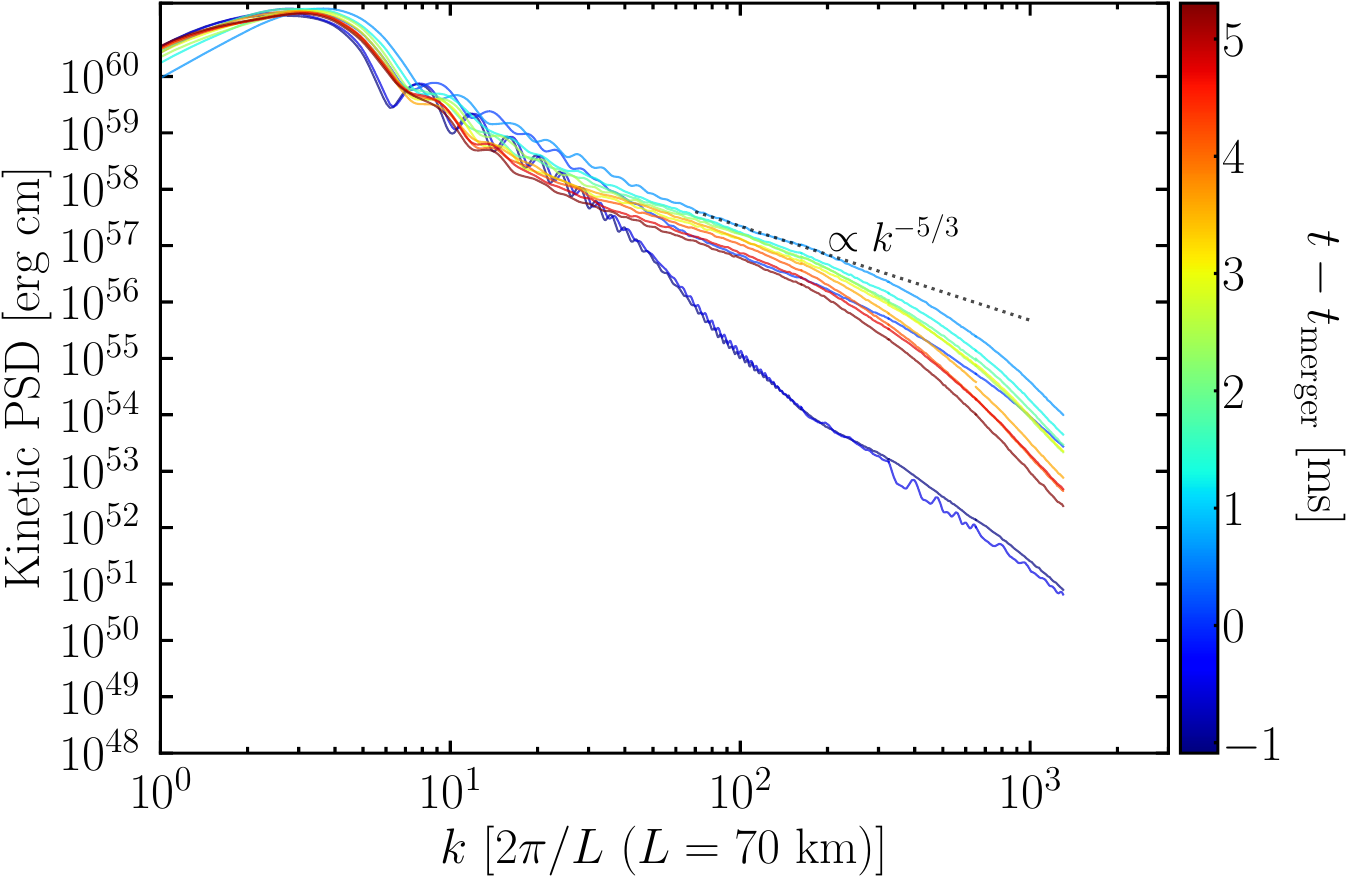}      
 	 \caption{ The same as Fig.~3 in the main paper, but with $\Delta x_{(16)}=12.5$~m.
      }\label{fig:PSD_low}
\end{figure*}

\subsubsection{Resolution study}
Figure~\ref{fig:PSD_low} plots the magnetic and kinetic PSD for the low resolution with $\Delta x_{(16)}=12.5$~m.

\bibliography{reference}

\end{document}